\documentclass[twocolumn,superscriptaddress]{revtex4-1}

\usepackage{graphicx}
\usepackage{amsmath}
\usepackage{amssymb}
\usepackage{setspace}
\usepackage{enumerate}
\usepackage{bm}
\usepackage{array}
\newcolumntype{C}[1]{>{\centering\let\newline\\\arraybackslash\hspace{0pt}}m{#1}}
\usepackage{microtype}
\usepackage{hyperref,color}
\usepackage{changes}

\usepackage{microtype}
\usepackage{multirow}

\usepackage{float}

\usepackage{float}
\usepackage{amsbsy}
\usepackage{xcolor, ulem,url}

\begin{document}
\title{Quasicrystal kirigami}

\author{Lucy Liu}
\thanks{These authors contributed equally to this work.}
\affiliation{Harvard College, Cambridge, MA 02138, USA}
\author{Gary P. T. Choi}
\thanks{These authors contributed equally to this work.}
\affiliation{Department of Mathematics, Massachusetts Institute of Technology, Cambridge, MA 02139, USA}
\author{L. Mahadevan}
\email{lmahadev@g.harvard.edu}
\affiliation{School of Engineering and Applied Sciences, Harvard University, Cambridge, MA 02138, USA}
\affiliation{Departments of Physics, and Organismic and Evolutionary Biology, Harvard University, Cambridge, MA 02138, USA}

\date{\today}

\begin{abstract}
Kirigami, the art of introducing cuts in thin sheets to enable articulation and deployment, has become an inspiration for a novel class of mechanical metamaterials with unusual properties. Here we complement the use of periodic tiling patterns for kirigami designs by showing that quasicrystals can also serve as the basis for designing deployable kirigami structures, and analyze the geometrical, topological and mechanical properties of these aperiodic kirigami structures.
\end{abstract}

\maketitle

\section{Introduction}
Kirigami is a traditional Japanese paper crafting art that has recently become popular among scientists and engineers. The simple idea of introducing cuts in a sheet of material has led to a surprisingly wide range of applications, including the design of super-stretchable materials~\cite{tang2017design}, nanocomposites~\cite{blees2015graphene,shyu2015kirigami}, energy-storing devices~\cite{song2015kirigami} and robotics~\cite{rafsanjani2018kirigami}. Numerous works have been devoted to the design of deployable kirigami patterns based on triangles~\cite{grima2006auxetic}, quads~\cite{grima2000auxetic,attard2008auxetic} or even ancient Islamic tiling patterns~\cite{rafsanjani2016bistable}, with recent efforts on generalizing their cut geometry~\cite{konakovic2016beyond,celli2018shape,konakovic2018rapid,choi2019programming,choi2020compact} and cut topology~\cite{chen2020deterministic,choi2020control,bossart2021oligomodal}. 

Almost without exception, these prior studies have manipulated the geometry, topology, and mechanics of tiling patterns with translational symmetry, most recently using periodic deployable kirigami patterns based on wallpaper groups~\cite{liu2021wallpaper}. However, the crystallographic restriction theorem states that the order of the rotational symmetry in periodic 2D patterns can only be 1, 2, 3, 4, or 6~\cite{grunbaum1986tilings}. This significantly limits the design space of periodic kirigami patterns. It is therefore natural to ask if kirigami based on patterns that lack translational or rotational symmetry might be possible. \textit{Quasicrystals}~\cite{shechtman1984metallic,levine1984quasicrystals,levine1986quasicrystals,socolar1986quasicrystals} and their tilings~\cite{wang1987two,socolar1989simple,baake1990ideal,baake1994classification,senechal1996quasicrystals,nagaoka2018single,ahn2018dirac} are a natural class of aperiodic structures that fit this bill, with three representative examples being the Penrose tiling~\cite{penrose1974role} (with 5-fold rotational symmetry), the Ammann--Beenker tiling~\cite{ammann1992aperiodic} (with 8-fold rotational symmetry), and the Stampfli tiling~\cite{stampfli1986dodecagonal} (with 12-fold rotational symmetry). Here we pose the problem of kirigami design from a new perspective: Is it possible to design radially deployable structures~\cite{you1997foldable,patel2007kinematic,kiper2008family,cabras2014auxetic} based on quasicrystal patterns? We solve this problem by proposing three different design methods and analyzing their geometrical, topological and mechanical properties. 

\section{Deployable quasicrystal design}
Our starting point is an aperiodic quasicrystal tiling pattern, which we seek to make deployable by cutting it along appropriate edges to articulate the structure while keeping it as a single connected whole. Here we show that we can achieve deployable \textit{symmetry-preserving} patterns, with the special quasicrystal rotation orders preserved upon deployment in all three approaches. Moreover, we focus on the design of \textit{rigid-deployable} quasicrystal patterns, in which all tiles do not undergo any bending or shearing throughout deployment. 

A helpful way to think about a kirigami pattern's structure is to consider its lattice representation, a graph where each tile is represented by a node. An edge between two nodes exists if their corresponding tiles are connected by a shared vertex around which both tiles can rotate freely. A pattern is rigidly deployable if it can be pulled apart along cuts so that tiles rotate away from each other and the pattern's enclosed area increases without compromising tiles' rigidity. As discussed in~\cite{liu2021wallpaper}, 3-cycles in the lattice representation of a pattern cannot be rigidly deployed. 

Consider three tiles lying so that any two tiles share an edge, like three regular hexagons that meet at a common vertex. If the tiles are connected in a 3-cycle, each pair of tiles must be connected at one end of their shared edge. No two tiles are able to rotate away from each other and deploy, because both tiles are also connected to the third tile, which is rigid and cannot accommodate any deformation. However, if we have four or more tiles connected in a cycle, connections can be designed so that when two connected tiles rotate away from each other, the other tiles rotate as well to accommodate the deployment while still satisfying the system's constraints. With this idea in mind, we consider changing the lattice connectivity of any given closed and compact tilings by (i) adding tiles, (ii) removing tiles, or (iii) directly changing the lattice connectivity without changing the number of tiles, so that the resulting lattice has no 3-cycles.

\begin{figure*}[t!]
    \centering
    \includegraphics[width=\textwidth]{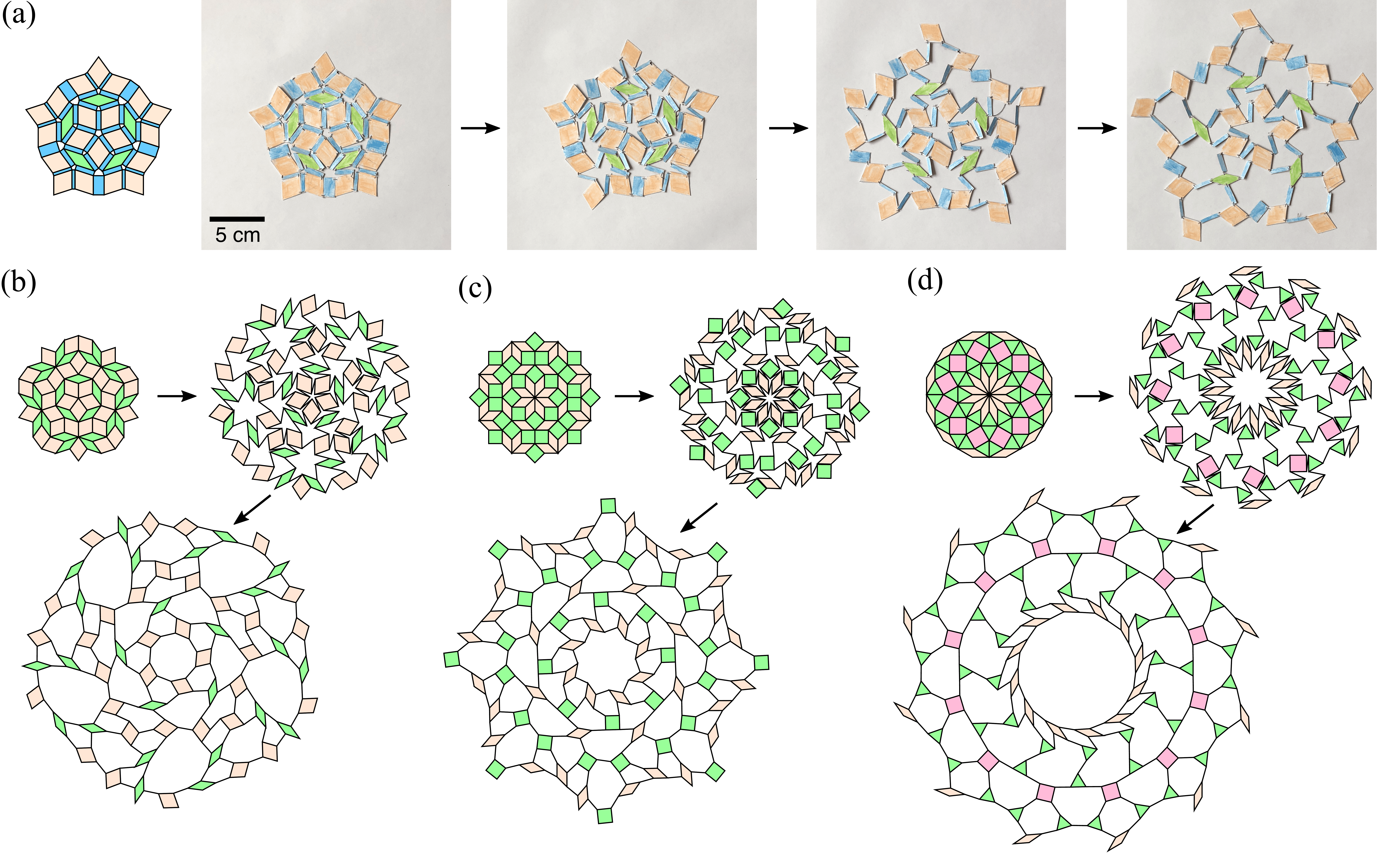}
    \caption{\textbf{Deployable quasicrystal patterns created using the expansion tile method}. (a) An example of the symmetry-preserving expansion method applied to the Penrose tiling and the deployment snapshots of a rigid cardstock paper model. (b) A deployable 5-fold Penrose tiling with ideal expansion tiles. (c) A deployable 8-fold Ammann--Beenker tiling with ideal expansion tiles. (d) A deployable 12-fold Stampfli tiling with ideal expansion tiles. For each example, the contracted state, an intermediate deployed state and the fully deployed state are shown.} 
    \label{fig:F1}
\end{figure*}

\begin{figure*}[t!]
    \centering
    \includegraphics[width=0.8\textwidth]{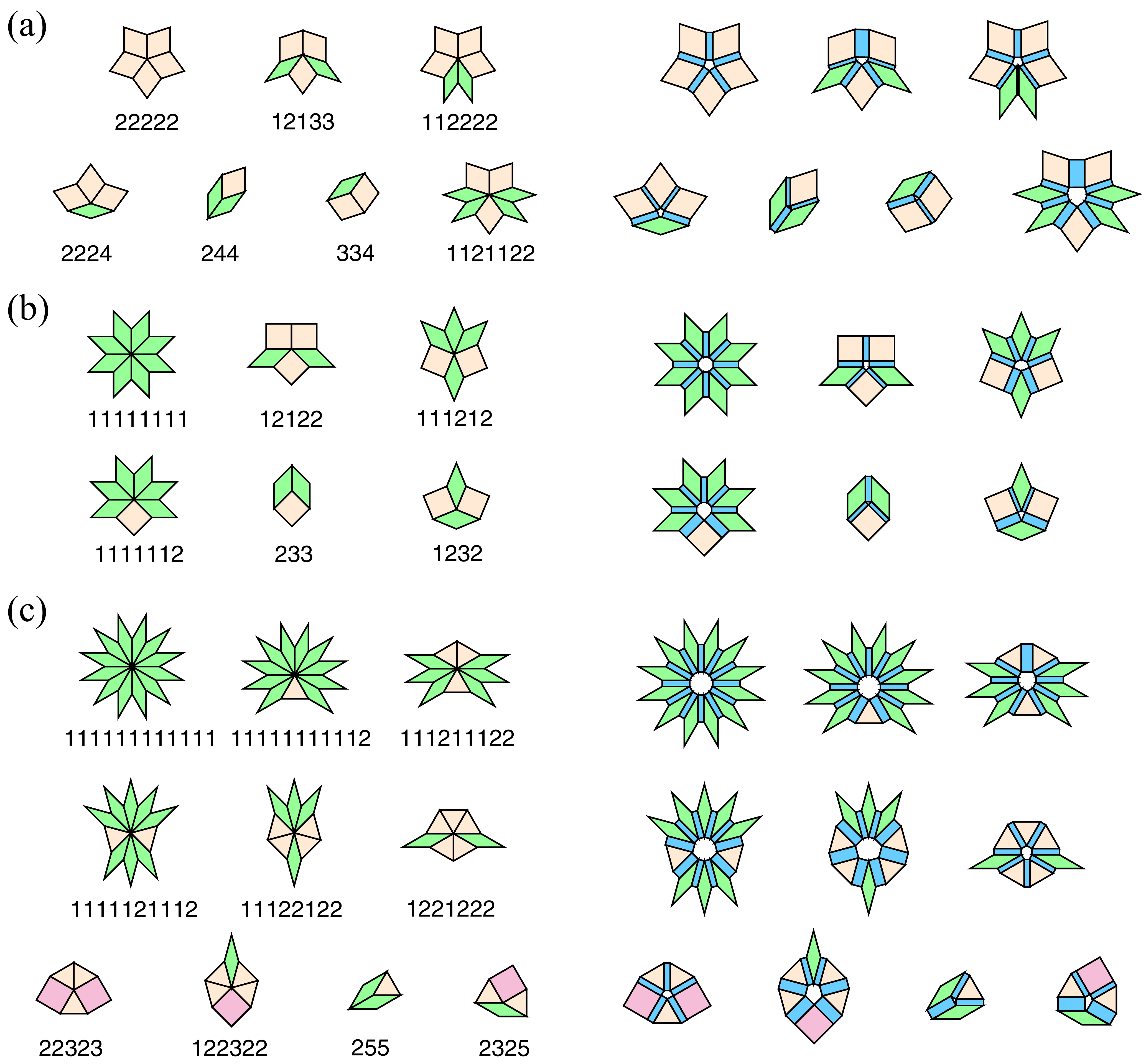}
    \caption{\textbf{The augmented version of the vertex stars using the expansion method.} (a) The Penrose vertex stars~\cite{senechal1996quasicrystals}. (b) The Ammann-Beenker vertex stars~\cite{baake1990ideal}. (c) The Stampfli vertex stars~\cite{baake1994classification}. The left column shows the original vertex stars, and the right column shows the augmented version of them with the expansion tiles colored in blue.}
    \label{fig:SI_vertex_star}
\end{figure*}

\begin{figure*}[t!]
    \centering
    \includegraphics[width=0.75\textwidth]{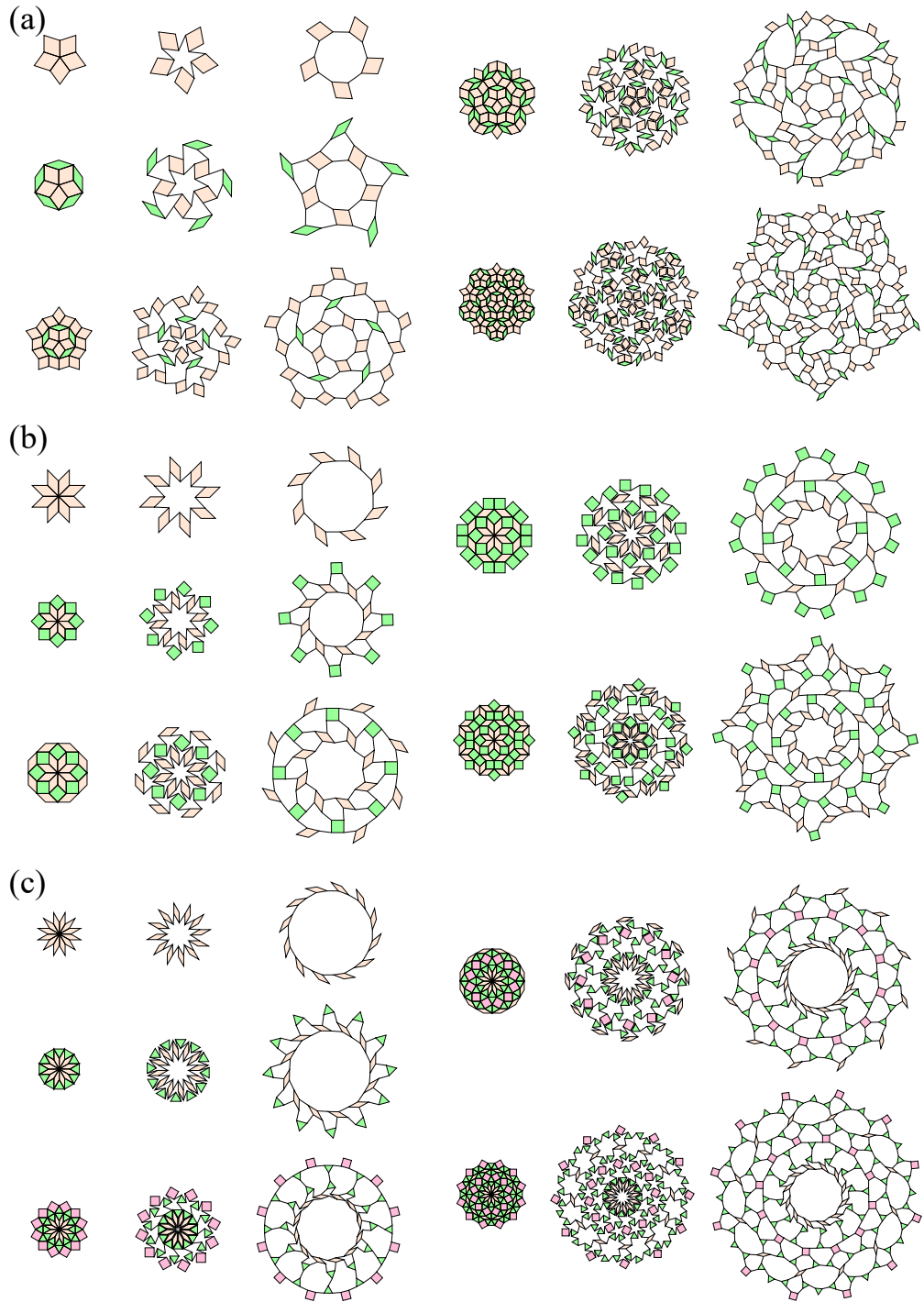}
    \caption{\textbf{Examples of deployable quasicrystal patterns produced using the expansion method.} (a) The Penrose patterns consist of 10, 25, 65, 165, 310 tiles (including the ideal expansion tiles) respectively. (b) The Ammann--Beenker patterns consist of 16, 40, 64, 104, 200 tiles (including the ideal expansion tiles) respectively. (c) The Stampfli patterns consist of 24, 60, 120, 216, 336 tiles (including the ideal expansion tiles) respectively. For each example, the contracted state, an intermediate deployed state and the fully deployed state are shown.}
    \label{fig:SI_expansion}
\end{figure*}

\subsection{The expansion tile method}
Our first approach for designing deployable quasicrystal patterns is to make use of the expansion tiles introduced in~\cite{liu2021wallpaper}, where thin tiles are added between existing tiles in the quasicrystal pattern. The new expansion tiles are connected to the tiles they are placed between, and they appear in the lattice representation as additional nodes in the middle of existing edges. Each expansion tile can also be considered as a new tile formed by introducing an extra cut on one of the two existing tiles near the edge shared by the two tiles. The 3-cycles in the lattice structure are turned into 6-cycles instead, and hence the entire pattern becomes deployable. 

To illustrate this idea, we fabricate a physical model of a deployable 5-fold Penrose pattern obtained by this method (see Fig.~\ref{fig:F1}(a) and Video~S1 of the Supplemental Material~\cite{supplementary}), which consists of rigid cardstock paper tiles connected by threads (see Appendix~\ref{sect:SI_physical} for more details). Note that the expansion tiles are not necessarily of the same width, and there may be gaps between the tiles in the pattern. To yield a closed and compact shape without gaps, one can consider ideal expansion tiles of infinitesimal width. Fig.~\ref{fig:F1}(b)--(d) show the simulated deployments of three deployable quasicrystal patterns with ideal expansion tiles (see also Video~S2--S4~\cite{supplementary}). It can be observed that the three patterns exhibit $5$, $8$ and $12$-fold symmetry throughout deployment from a closed and compact contracted configuration to the fully deployed configuration, and a large size change is achieved. Here, the deployment simulations are performed using Python, with the 2D rigid body physics library Pymunk utilized. The deployment of each pattern is modeled by continually applying forces on the pattern's convex hull tiles, in the direction away from the pattern center (see Appendix~\ref{sect:SI_simulation} for more details). 

To explain the idea more systematically, Fig.~\ref{fig:SI_vertex_star}(a) shows the augmented version of the seven types of Penrose vertices~\cite{senechal1996quasicrystals} using the expansion method. Given a Penrose tiling of any size, we can consider it as a combination of the seven vertex stars and augment the tiling accordingly, thereby producing a deployable Penrose pattern. Similarly, one can augment the Ammann-Beenker vertex stars~\cite{baake1990ideal} (Fig.~\ref{fig:SI_vertex_star}(b)) and the Stampfli vertex stars~\cite{baake1994classification} (Fig.~\ref{fig:SI_vertex_star}(c)) to make them deployable. We describe each vertex star using the ratios of the tile angles meeting at the center of the star. For instance, Penrose vertex star 22222 has five $72$-degree angles meeting at the center, while Penrose vertex star 12133 has angles of 36, 72, 36, 108, and 108 degrees meeting at the center. One can further eliminate the gaps in between the tiles by considering ideal expansion tiles with infinitesimal width.

Fig.~\ref{fig:SI_expansion}(a) shows several examples of deployable Penrose patterns with ideal expansion tiles produced using this method. Similarly, one can augment an 8-fold Ammann--Beenker tiling of any size using the expansion tiles and produce a deployable pattern. Fig.~\ref{fig:SI_expansion}(b) shows several examples of deployable Ammann--Beenker patterns produced using this method. Fig.~\ref{fig:SI_expansion}(c) shows several examples of deployable 12-fold Stampfli patterns produced using this method. It can be observed that the rotational symmetry of the quasicrystal patterns is preserved throughout the deployment.

\begin{figure*}[t!]
    \centering
    \includegraphics[width=0.65\textwidth]{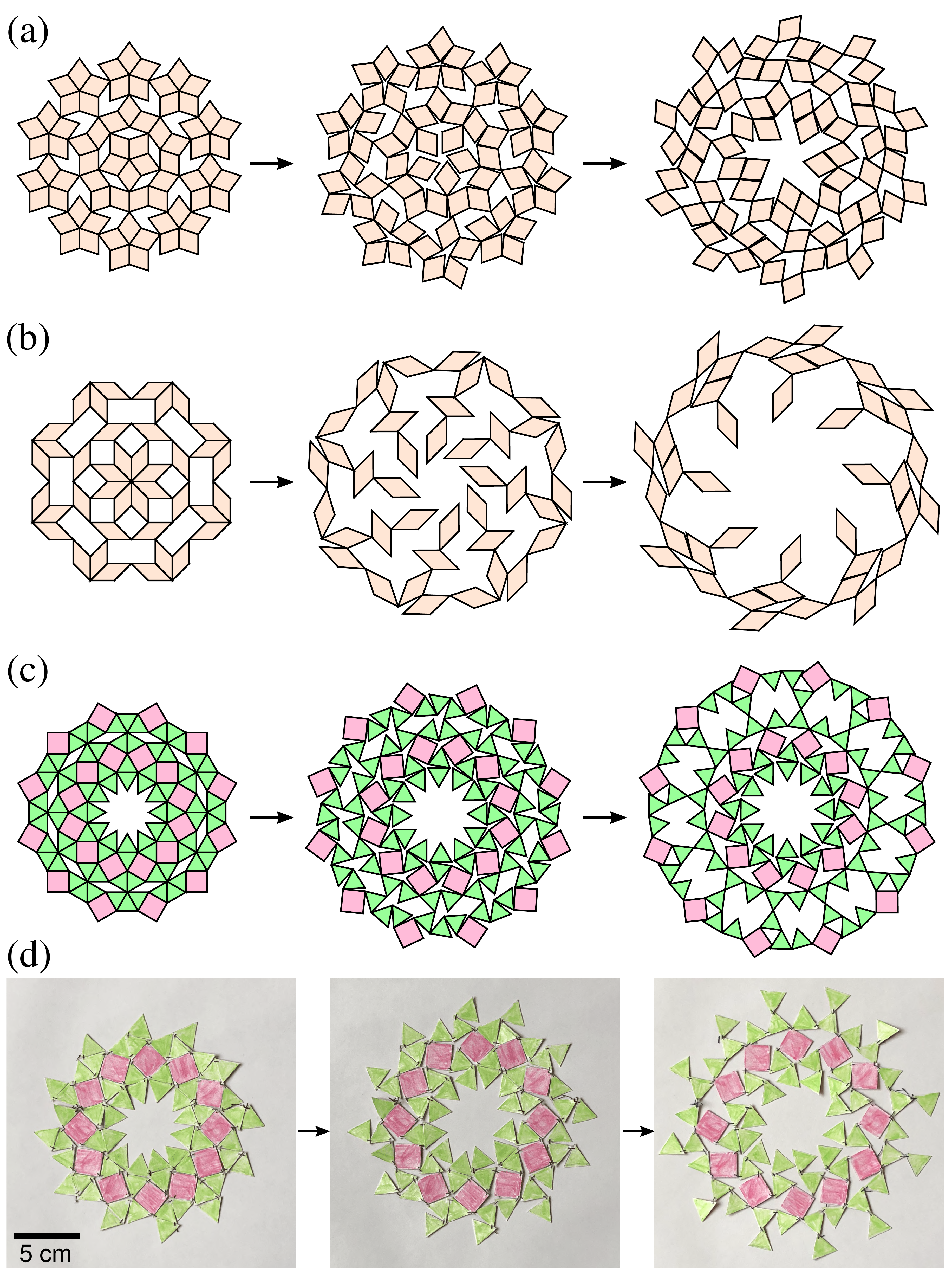}
    \caption{\textbf{Deployable quasicrystal patterns created using the tile removal method}. By removing certain tiles in a given quasicrystal pattern, we create holes and hence allow the pattern to be deployed. (a)~A deployable 5-fold Penrose tiling. (b)~A deployable 8-fold Ammann--Beenker tiling. (c)~A deployable 12-fold Stampfli tiling. For each example, the contracted state, an intermediate deployed state and the fully deployed state are shown. (d)~The deployment snapshots of a rigid cardstock paper model of a deployable Stampfli pattern.}
    \label{fig:F2}
\end{figure*}

\begin{figure*}[t!]
    \centering
    \includegraphics[width=0.9\textwidth]{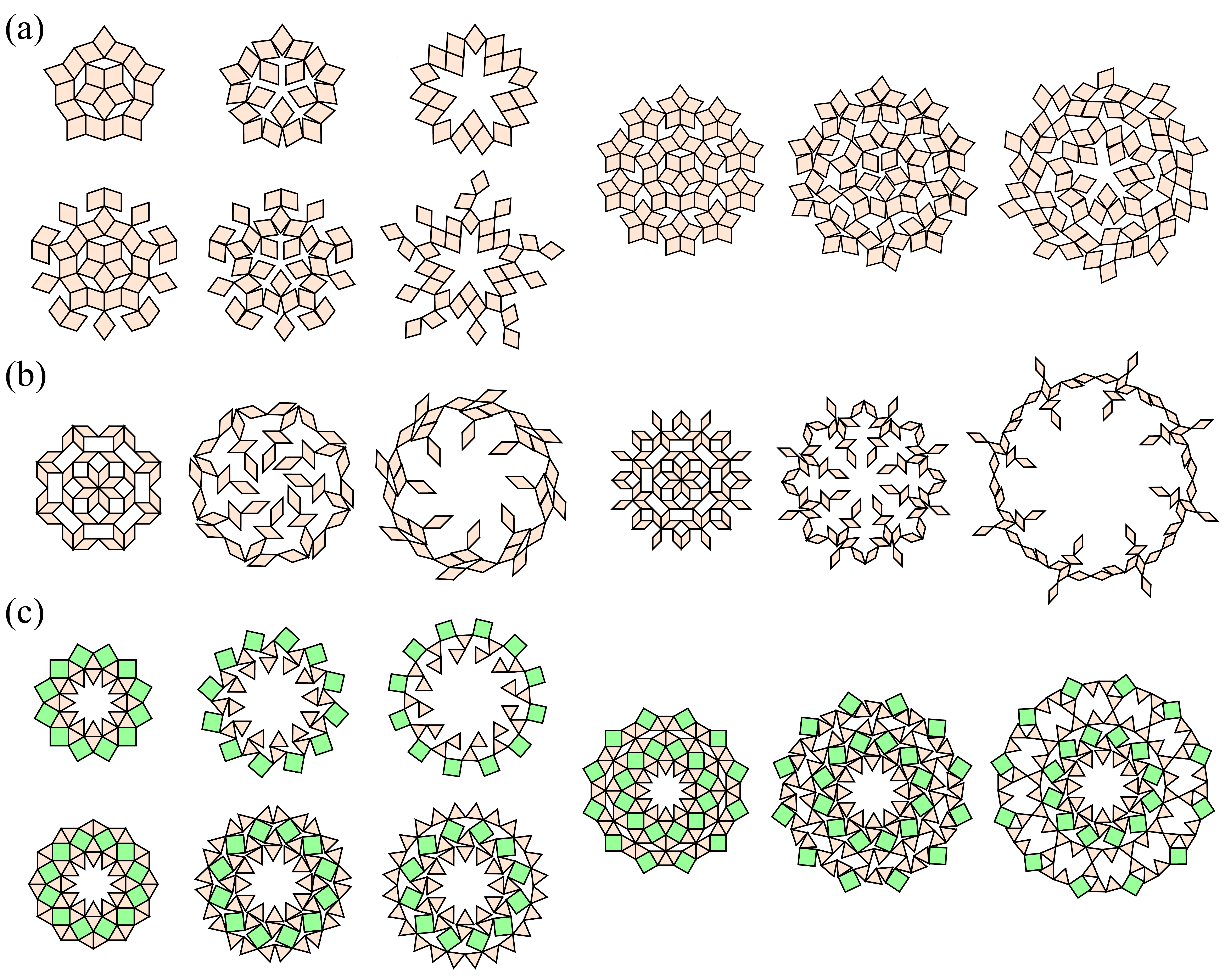}
    \caption{\textbf{Examples of deployable quasicrystal patterns produced using the removal method.} (a) The Penrose patterns consist of 20, 35 and 70 tiles respectively. (b) The Ammann--Beenker patterns consist of 40 and 64 tiles respectively. (c) The Stampfli patterns consist of 36, 60 and 108 tiles respectively. For each example, the contracted state, an intermediate deployed state and the fully deployed state are shown.}
    \label{fig:SI_removal}
\end{figure*}

\subsection{The tile removal method}
Our second approach for achieving deployability is removing tiles from a given quasicrystal pattern, changing the lattice connectivity and introducing negative space. By taking a tile involved in each 3-cycle out of the pattern, we can again remove 3-cycles in the lattice and make the structure deployable. 

For instance, a deployable 5-fold Penrose pattern can be obtained by removing one type of rhombus tile in the tiling (Fig.~\ref{fig:F2}(a)). Similarly, a deployable 8-fold Ammann--Beenker pattern can be obtained by removing all squares (Fig.~\ref{fig:F2}(b)), and a deployable 12-fold Stampfli pattern can be obtained by removing all rhombi (Fig.~\ref{fig:F2}(c)). Analogous to the expansion method, the deployable patterns produced by the tile removal method exhibit 5, 8 and 12-fold symmetry throughout deployment (see also Video S5--S7~\cite{supplementary}). Fig.~\ref{fig:F2}(d) shows a physical model of a deployable Stampfli pattern (see also Video~S8~\cite{supplementary}). Fig.~\ref{fig:SI_removal} shows more examples of deployable Penrose, Ammann--Beenker and Stampfli patterns produced using this method, from which it can again be observed that the rotational symmetry is preserved throughout the deployment. 

We remark that the tile removal method only works for patterns with a sufficiently large number of tiles. For instance, if we only consider the five innermost tiles of the Penrose tiling, it is impossible to remove certain tiles without breaking the symmetry. Also, while this method does not achieve a large size change because of the holes, it is useful for applications that require changing the size and shape of the holes throughout deployment without changing the size of the entire structure much. For instance, one may design a flexible filter that allows some shapes to pass through at the initial state, and some other shapes to pass through at the deployed state.

\begin{figure*}[t!]
    \centering
    \includegraphics[width=0.9\textwidth]{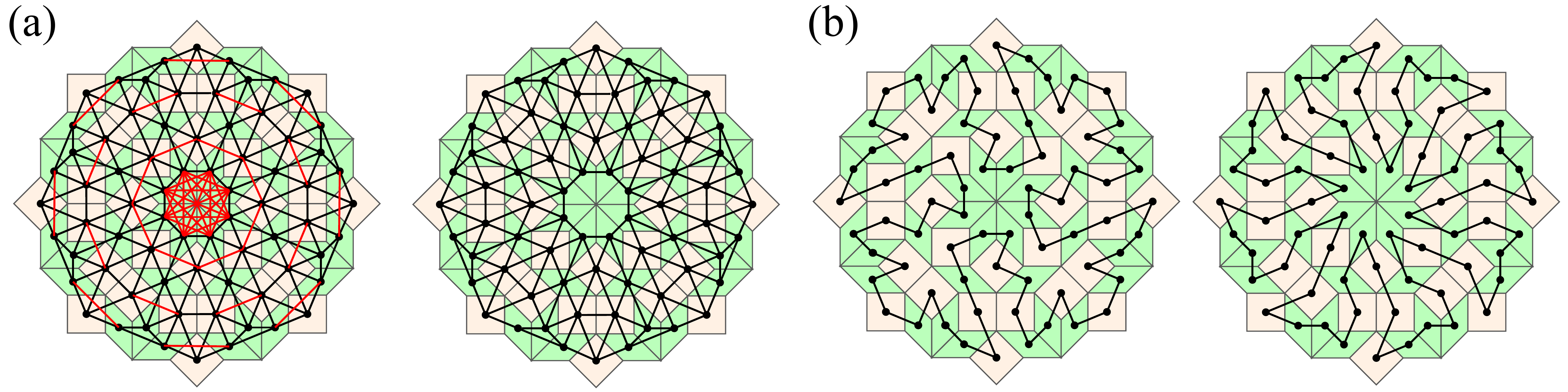}
    \caption{\textbf{Finding Hamiltonian cycles in a quasicrystal pattern.} (a) Given an Ammann--Beenker pattern, we first consider the underlying graph $G$ (left), which is 4-connected but may contain some edge crossings. By removing the red edges, we obtain a 4-connected planar subgraph $\widetilde{G}$ (right). (b) Two different Hamiltonian cycles extracted from the subgraph $\widetilde{G}$ in (a).}
    \label{fig:SI_hamiltonian_cycle}
\end{figure*}

\begin{figure*}[t!]
    \centering
    \includegraphics[width=\textwidth]{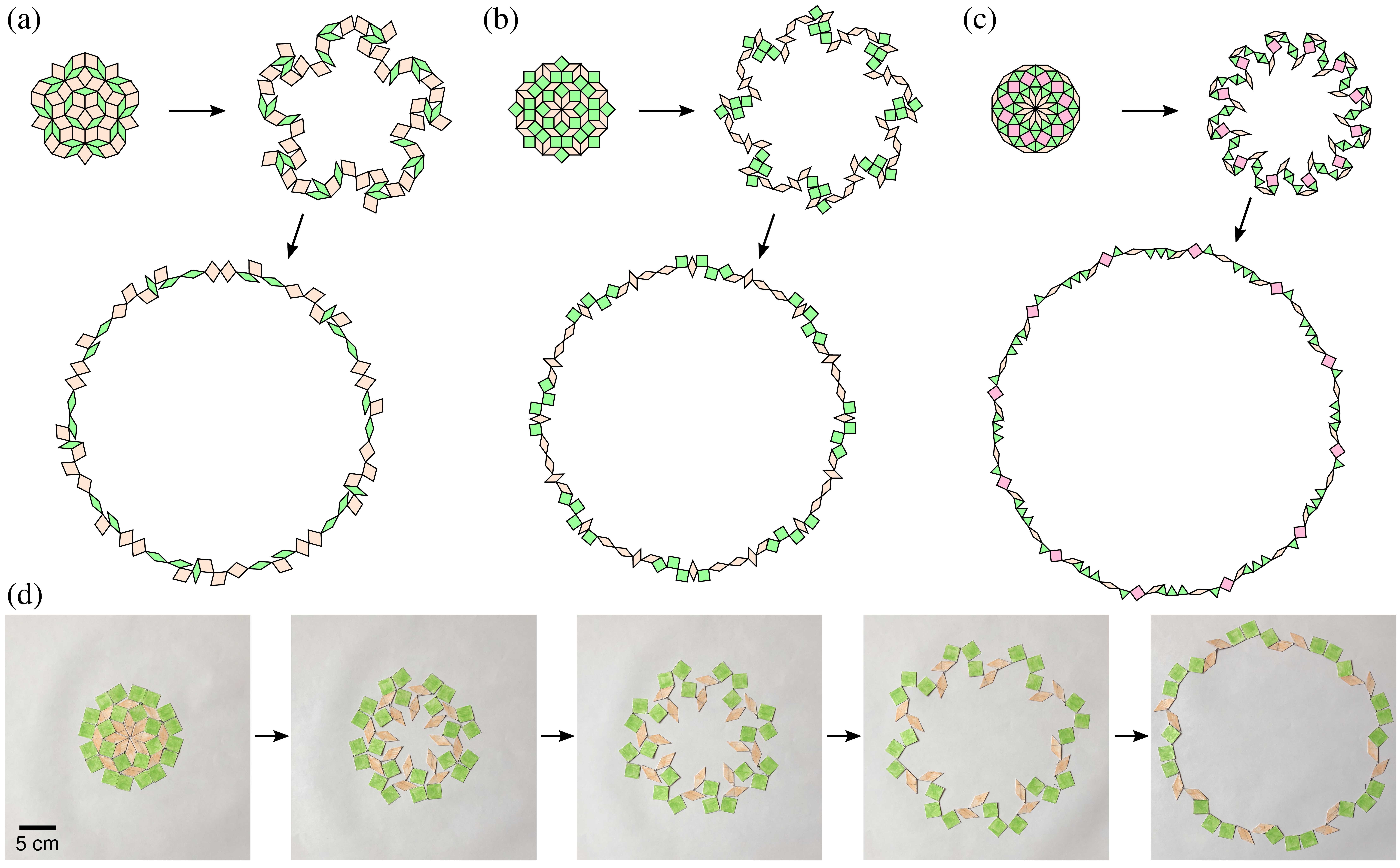}
    \caption{\textbf{Deployable quasicrystal patterns created using the Hamiltonian cycle method}. We start by constructing a planar subgraph of the connectivity graph of the given quasicrystal pattern. We can then find a Hamiltonian cycle in the planar subgraph, which passes through all tiles exactly once and gives us a deployable structure. (a)~A deployable 5-fold Penrose tiling. (b)~A deployable 8-fold Ammann--Beenker tiling. (c)~A deployable 12-fold Stampfli tiling. For each example, the contracted state, an intermediate deployed state and the fully deployed state are shown. (d)~The deployment snapshots of a rigid cardstock paper model of a deployable Ammann--Beenker pattern.}
    \label{fig:F3}
\end{figure*}

\begin{figure*}[t!]
    \centering
    \includegraphics[width=0.78\textwidth]{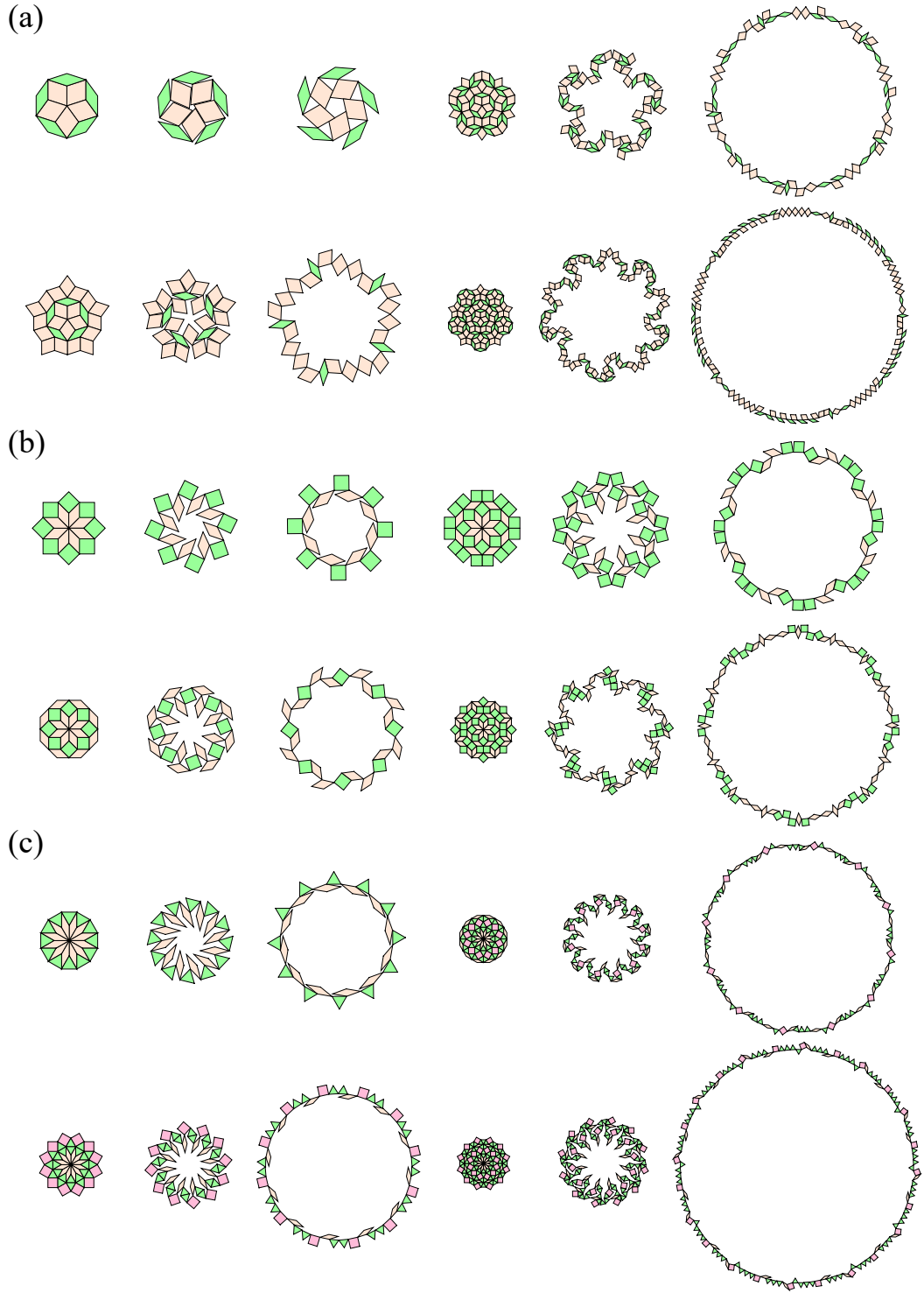}
    \caption{\textbf{Examples of deployable quasicrystal patterns produced using the Hamiltonian method.} (a) The Penrose patterns consist of 10, 25, 60, 110 tiles respectively. (b) The Ammann--Beenker patterns consist of 8, 16, 24, 40, 72 tiles respectively. (c) The Stampfli patterns consist of 12, 24, 48, 84, 132 tiles respectively. For each example, the contracted state, an intermediate deployed state and the fully deployed state are shown.}
    \label{fig:SI_hamiltonian}
\end{figure*}

\begin{figure}[t!]
    \centering
    \includegraphics[width=0.45\textwidth]{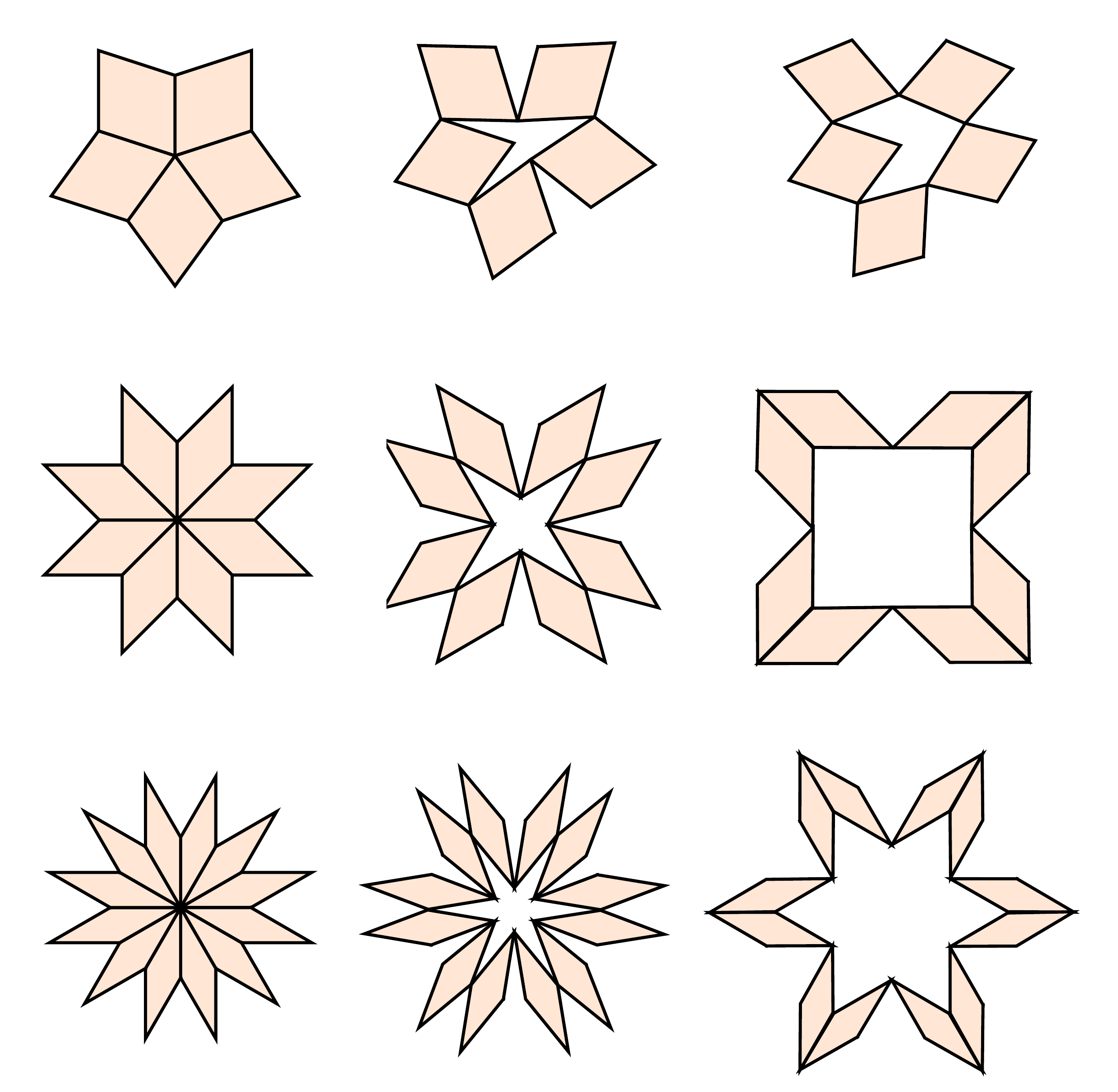}
    \caption{\textbf{Examples of deployable quasicrystal patterns produced using the Hamiltonian method with symmetry lost under deployment.} The 5-fold Penrose pattern with 5 tiles becomes 1-fold under deployment, the 8-fold Ammann--Beenker pattern with 8 tiles becomes 4-fold under deployment, and the 12-fold Stampfli pattern with 12 tiles becomes 6-fold under deployment.}
    \label{fig:SI_hamiltonian_symmetry_lost}
\end{figure}

\subsection{The Hamiltonian cycle method}
Our third method which does not require us to add or remove any tiles is based on manipulating the connectivity of the tiles. Furthermore, it is possible in edge to edge polygonal tilings to optimize expansion by connecting the tiles in a Hamiltonian cycle, which deploys into a single loop of connected tiles. We introduce the following graphic-theoretic approach to achieve this. 

Consider the lattice representation of a pattern, i.e. a graph $G$ where the nodes are the tile centers and there exists an edge between two nodes if and only if the two corresponding tiles share a connected vertex~\cite{liu2021wallpaper} (Fig.~\ref{fig:SI_hamiltonian_cycle}(a), left). By the Tutte theorem~\cite{tutte1956theorem}, every 4-connected planar graph has a Hamiltonian cycle (i.e. a closed loop that visits all nodes exactly once). For all three quasicrystal patterns we consider, note that each tile has at least three sides and there are always some other tiles that share a common vertex with it, making the vertex degree $\geq 4$ for all nodes in the associated graph $G$. Although $G$ may not be planar (i.e. there may be edge crossings), one can always consider a subgraph $\widetilde{G}$ of $G$ with a few edges connecting tiles in the same layer removed, thereby avoiding edge crossings while keeping the vertex degree $\geq 4$. For instance, by removing the edges highlighted in red, we obtain a 4-connected planar subgraph $\widetilde{G}$ (Fig.~\ref{fig:SI_hamiltonian_cycle}(a), right). Consequently, based on the 4-connected planar subgraph $\widetilde{G}$, we can draw a Hamiltonian cycle and hence obtain a deployable structure with all tiles used. It is noteworthy that such Hamiltonian cycles are not necessarily unique. Fig.~\ref{fig:SI_hamiltonian_cycle}(b) shows two different Hamiltonian cycles, which lead to two different deployable Ammann--Beenker patterns.

Fig.~\ref{fig:F3}(a)--(c) show three examples of deployable Penrose, Ammann--Beenker and Stampfli patterns obtained by this method, in which a significant size change can be observed throughout the symmetry-preserving deployment (see Video~S9--S11~\cite{supplementary}). A physical model of a deployable Ammann--Beenker pattern is shown in Fig.~\ref{fig:F3}(d) (see also Video~S12~\cite{supplementary}). Fig.~\ref{fig:SI_hamiltonian} shows more examples produced using the Hamiltonian method, with their rotational symmetry preserved throughout the deployment. We remark that if the number of tiles is too small, the resulting deployable structures may be with symmetry lost under the deployment (see Fig.~\ref{fig:SI_hamiltonian_symmetry_lost}).

It is natural to consider the problem of finding the largest Hamiltonian cycle, which can be thought of as a traveling salesman problem. Each tile in the Hamiltonian cycle is connected to other tiles at exactly two of its vertices. We can consider trying to maximize the sum $\sum_{i} \text{dist}(a_i, b_i)$ for all tiles $i$, where $a_i$ and $b_i$ are the two vertices of tile $i$ that are constrained to vertices of other tiles. This is the length of the Hamiltonian path, which after deployment will become approximately the perimeter of the deployed pattern. It is maximized by the longest path that starts at a vertex of one tile and ``travels'' through all other tiles, entering and exiting each tile via different vertices. The distances between the entrance and exit vertices on each tile comprise the lengths that make up the final path length. Rotationally symmetric cycles can be found by considering the Hamiltonian path on a rotational symmetry unit (e.g. one fifth of the Penrose tiling or one eighth of the Ammann-Beenker tiling) that starts and ends at two vertices which would be adjacent to each other in the full pattern. However, naive dynamic programming for this problem fails to account for the fact that edge crossings between pairs of tiles at the same vertex star will cause a Hamiltonian path found via dynamic programming to be ``twisted" and undeployable in two dimensions. 

\begin{figure*}[t]
    \centering
    \includegraphics[width=\textwidth]{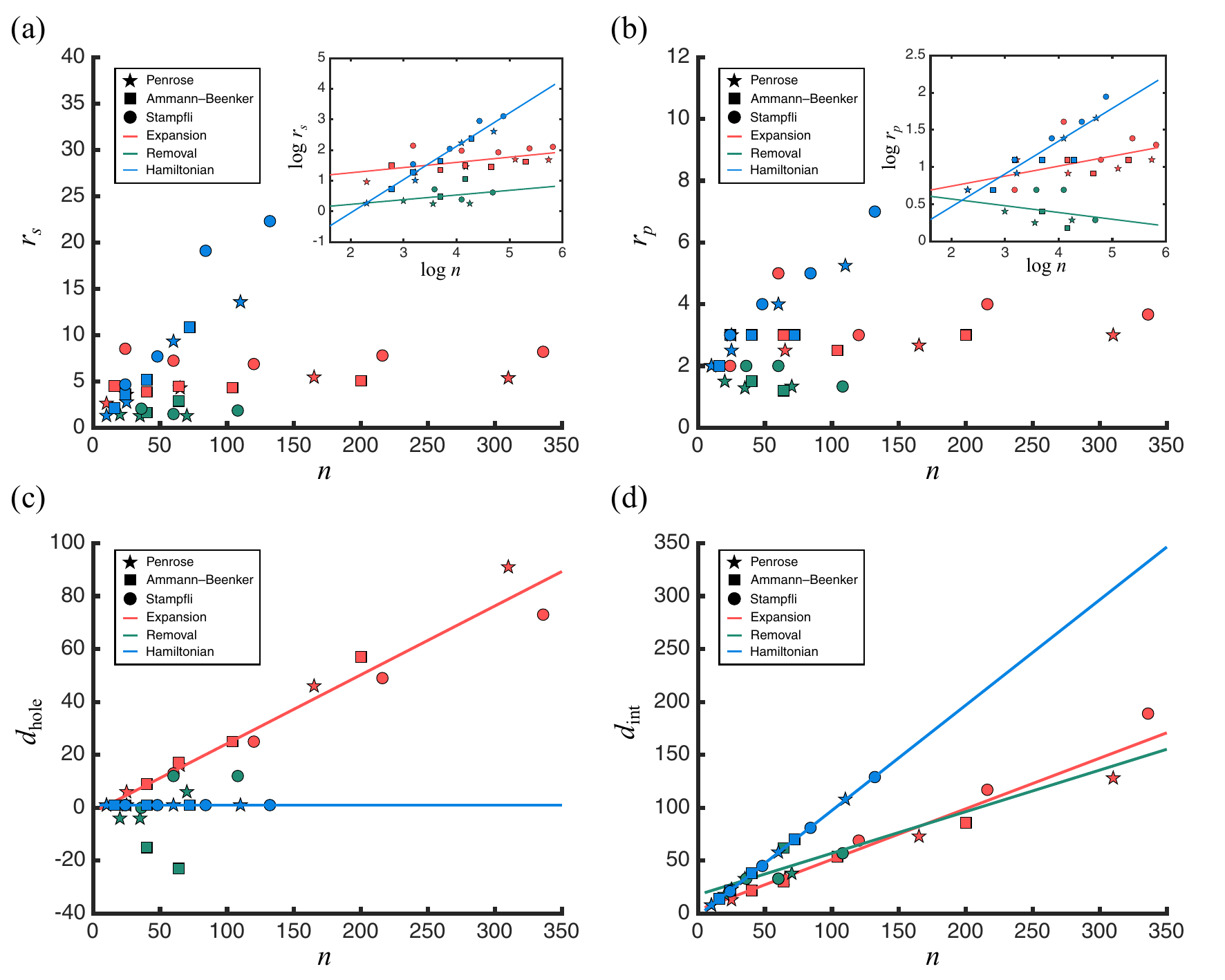}
    \caption{\textbf{Geometric, topological and mechanical properties of the deployable quasicrystal patterns}. (a)~The size change ratio $r_s$ of the deployable Penrose, Ammann--Beenker and Stampfli patterns produced by the three proposed methods versus the number of tiles $n$. Inset shows the log-log plot. (b) The perimeter change ratio $r_p$ of the patterns. Inset shows the log-log plot. (c) The change in the number of holes $d_{\text{hole}}$ under deployment. (d) The internal degree of freedom $d_{\text{int}}$ of the patterns. For each plot, the three different types of patterns are represented by three different types of markers, and the three different methods are represented by three different colors.}
    \label{fig:SI_F4}
\end{figure*}

\section{Geometrical, topological and mechanical properties}
After establishing the three above design methods for producing deployable and symmetry-preserving quasicrystal patterns, it is natural to ask how the patterns produced by the three methods differ in terms of their geometrical, topological and mechanical properties. 

\subsection{Geometry}
To study the geometric properties of the deployable quasicrystal patterns, it is natural to consider the change in size and perimeter of them under the deployment. Here we define the size change ratio (SCR) by 
\begin{equation}
    r_s = \frac{\text{Area of the fully deployed pattern}}{\text{Area of the contracted pattern}}
\end{equation}
and the perimeter change ratio (PCR) by
\begin{equation}
r_p = \frac{\text{Perimeter of the fully deployed pattern}}{\text{Perimeter of the contracted pattern}}. 
\end{equation}
Table~\ref{table:SI_scr_vertex_star} records the SCR $r_s$ of different vertex stars in the deployable Penrose, Ammann--Beenker, and Stampfli patterns obtained by the expansion method, from which it can be observed that most vertex stars in a given pattern have a comparable $r_s$. Therefore, while we consider a radial deployment of the patterns, we can achieve a largely uniform deployment effect in the final deployed shape. Table~\ref{table:scr} and Table~\ref{table:pcr} record the SCR and the PCR of the deployable Penrose, Ammann--Beenker, and Stampfli patterns with different resolution produced by the three design methods. Fig.~\ref{fig:SI_F4}(a)--(b) show the SCR and PCR plots for the deployable Penrose, Ammann--Beenker and Stampfli patterns produced by the three proposed methods.

\begin{table}[t!]
\small
    \centering
    \begin{tabular}{c|c|c}
        \textbf{Pattern} & \textbf{Vertex star type} & \ \ $r_s$ \ \  \\ \hline
                  & 22222 & 2.61 \\				
	                                                & 12133 & 2.79 \\	
                                                    & 112222 & 3.21 \\				
        Penrose	(5-fold)                                        & 2224 & 2.35 \\
                                                   & 244 & 2.08 \\ 
                                                    & 334 & 1.99 \\ 
                                                    & 1121122 & 3.89 \\  \hline
         & 11111111 & 4.51 \\	
                                                    & 1111112 & 3.90 \\
       Ammann--Beenker                                             & 12122 & 2.71 \\
        (8-fold)                                            & 233 & 2.06 \\
                                                    & 111212 & 3.28 \\
                                                    & 1232 & 2.38 \\\hline
               & 111111111111 & 8.52 \\	
                                                    & 11111111112 & 7.97 \\
                                                    & 111211122 & 6.80 \\
                                                    & 1111121112 & 7.39 \\
       Stampfli                                             & 11122122 & 6.27 \\
        (12-fold)                                            & 1221222 & 5.70 \\
                                                    & 22323 & 3.32 \\
                                                    & 122322 & 4.40 \\
                                                    & 255 & 2.68 \\
                                                    & 2325 & 2.98 \\
    \end{tabular}
    \caption{\textbf{The size change ratio (SCR) $r_s$ of different vertex stars in the deployable Penrose, Ammann--Beenker, and Stampfli patterns produced by the expansion method.} Here we consider the expansion tiles to be of infinitesimal width.}
    \label{table:SI_scr_vertex_star}
\end{table}

\begin{table*}[t!]
\small
    \centering    
    \begin{tabular}{c|c|c|c|c|c|c|c}
        \multirow{2}{*}{\textbf{Pattern}} & \multirow{2}{*}{\textbf{\# layers}} & \multicolumn{2}{c|}{\textbf{Expansion}} & \multicolumn{2}{c|}{\textbf{Removal}} & \multicolumn{2}{c}{\textbf{Hamiltonian}}\\ \cline{3-8}
        & & \# tiles & $r_s$ & \# tiles & $r_s$ & \# tiles & $r_s$\\ \hline
       \multirow{5}{*}{Penrose (5-fold)}          & 1 & 10 & 2.62 & / & / & / & /\\	
	                                              & 2 & 25 & 3.58 & / & / & 10 & 1.30	\\	
                                                  & 3 & 65 & 4.28 & 20 & 1.41 & 25 & 2.75\\
        	                                      & 4 & 165 & 5.44   & 35 & 1.28 & 60 & 9.32\\
                                                  & 5 & 310 & 5.36  & 70 & 1.29 & 110 & 13.57 \\  \hline
        \multirow{5}{*}{Ammann--Beenker (8-fold)} & 1 & 16 & 4.51  & / & / & / & /\\ 
	                                              & 2 & 40 & 3.86 &  / & / & 16 & 2.09 \\ 
	                                              & 3 & 64 & 4.47 & / & / & 24 & 3.60\\ 
	                                              & 4 & 104 & 4.31 & 40 & 1.62 & 40 & 5.18\\ 
	                                              & 5 & 200 & 5.06 & 64 & 2.88 & 72 & 10.86\\	\hline
       \multirow{5}{*}{Stampfli (12-fold)}	      & 1 & 24 & 8.52 & / & / & / & /\\	 
                                                  & 2 & 60 & 7.23 &  / & / & 24 &  4.66\\	
                                        	      & 3 & 120 & 6.87 & 36 & 2.05 & 48 & 7.69\\
	                                              & 4 & 216 & 7.79 & 60 & 1.47 & 84 & 19.10 \\
	                                              & 5 & 336 & 8.19 & 108 & 1.85 & 132 & 22.30\\
    \end{tabular}
    \caption{\textbf{The size change ratio $r_s$ of the deployable Penrose, Ammann--Beenker, and Stampfli patterns produced by the three different design methods.}}
    \label{table:scr}
\end{table*}

\begin{table*}[t]
\small
    \centering    
    \begin{tabular}{c|c|c|c|c|c|c|c}
        \multirow{2}{*}{\textbf{Pattern}} & \multirow{2}{*}{\textbf{\# layers}} & \multicolumn{2}{c|}{\textbf{Expansion}} & \multicolumn{2}{c|}{\textbf{Removal}} & \multicolumn{2}{c}{\textbf{Hamiltonian}}\\ \cline{3-8}
        & & \# tiles & $r_p$ & \# tiles & $r_p$ & \# tiles & $r_p$\\ \hline
       \multirow{5}{*}{Penrose (5-fold)}          & 1 & 10 & 2 & / & / & / & /\\				
	                                              & 2 & 25 & 3 & / & / & 10 & 2	\\	
                                                  & 3 & 65 & $5/2$ & 20 & $3/2$ & 25 & $5/2$ \\
        	                                      & 4 & 165 & $8/3$ & 35 & $9/7$ & 60 & 4 \\
                                                  & 5 & 310 & 3 & 70 & $4/3$ & 110 & $21/4$ \\  \hline
        \multirow{5}{*}{Ammann--Beenker (8-fold)} & 1 & 16 & 2 & / & / & / & /\\
	                                              & 2 & 40 & 3 &  / & / & 16 & 2 \\ 
	                                              & 3 & 64 & 3 & / & / & 24 & 3 \\ 
	                                              & 4 & 104 & $5/2$ & 40 & $3/2$ & 40 & 3\\ 
	                                              & 5 & 200 & 3 & 64 & $6/5$ & 72 & 3 \\	\hline
       \multirow{5}{*}{Stampfli (12-fold)}	      & 1 & 24 & 2 & / & / & /  & /\\	
                                                  & 2 & 60 & 5 &  / & / & 24 & 3\\	
                                        	      & 3 & 120 & 3 & 36 & 2 & 48 & 4\\				
	                                              & 4 & 216 & 4 & 60 & 2 & 84 & 5 \\
	                                              & 5 & 336 & $11/3$ & 108 & $4/3$ & 132 & 7\\
    \end{tabular}
    \caption{\textbf{The perimeter change ratio $r_p$ of the deployable Penrose, Ammann--Beenker, and Stampfli patterns produced by the three different design methods.}}
    \label{table:pcr}
\end{table*}

Note that when the number of tiles $n$ is small (i.e. only the first few layers around the center of the pattern are used), the expansion method results in the largest SCR. As $n$ increases, the Hamiltonian method achieves the largest SCR and PCR among the three design methods. From the log-log plots of $r_s$ and $r_p$ (Fig.~\ref{fig:SI_F4}(a) inset and Fig.~\ref{fig:SI_F4}(b) inset), it can be observed that $r_s$ and $r_p$ increase with $n$ following the power law $r_s \propto n$ and $r_p \propto \sqrt{n}$ for the Hamiltonian method. 

To explain this, let $l_{\text{min}}$ and $l_{\text{max}}$ be the minimum and maximum length of the edges and the diagonals of the tiles respectively, $a_{\text{min}}$ and $a_{\text{max}}$ be the minimum and maximum area of the tiles respectively, and $A(n)$ be the area bounded by the fully deployed quasicrystal pattern. By the construction of the Hamiltonian method, $A(n)$ should be not less than the area bounded by the circle formed by the shortest edges of every tile. It should also not be greater than the area bounded by the circle formed by the longest edges of every tile, plus the sum of the areas of each tile (as the tiles may lie outside of the circle formed). Therefore, we have
\begin{equation}
    r_s(n) \geq \frac{A(n)}{na_{\text{max}}} \geq \frac{\pi (nl_{\text{min}}/2\pi)^2}{n a_{\text{max}}}= \frac{l_{\text{min}}^2}{4\pi a_{\text{max}}} n
\end{equation} 
and
\begin{equation}
    r_s(n) \leq \frac{A(n)}{na_{\text{min}}} \leq \frac{\pi \left(\frac{nl_{\text{max}}}{2\pi}\right)^2 + na_{\max}}{n a_{\text{min}}}= \frac{nl_{\text{max}}^2 + a_{\text{max}}}{4\pi a_{\text{min}}}.
\end{equation}
The two inequalities show that $r_s \propto n$. Furthermore, as $r_s$ is bounded below and above by some multiples of $n$, the radius change is bounded below and above by some other multiples of $\sqrt{n}$. This implies that
\begin{equation}
    k_1 \sqrt{n} \leq r_p(n) \leq k_2 \sqrt{n}
\end{equation}
for some constants $k_1, k_2$, and hence $r_p \propto \sqrt{n}$. 

\subsection{Topology}
We then study the topological property of the deployable quasicrystal patterns by assessing the change in the number of holes of them (denoted as $d_{\text{hole}}$) throughout deployment. Table~\ref{table:holes} records the change in $d_{\text{hole}}$ of the quasicrystal patterns throughout the deployment. Here, we only consider the holes which are directly related to the holes in the underlying connectivity graph of a pattern. In other words, the holes that are formed geometrically using some floppy tiles throughout the deployment (e.g. the outermost triangular holes in the last example of Fig.~\ref{fig:SI_removal}(c)) are not considered. Fig.~\ref{fig:SI_F4}(c) shows the plot of $d_{\text{hole}}$ for the deployable Penrose, Ammann--Beenker and Stampfli patterns produced by the three proposed methods.

\begin{table*}[t!]
\small
    \centering
    \begin{tabular}{c|c|c|c|c|c|c|c}
        \multirow{2}{*}{\textbf{Pattern}} & \multirow{2}{*}{\textbf{\# layers}} & \multicolumn{2}{c|}{\textbf{Expansion}} & \multicolumn{2}{c|}{\textbf{Removal}} & \multicolumn{2}{c}{\textbf{Hamiltonian}}\\ \cline{3-8}
        & & \# tiles & $d_{\text{hole}}$ & \# tiles & $d_{\text{hole}}$ & \# tiles & $d_{\text{hole}}$\\ \hline
       \multirow{5}{*}{Penrose (5-fold)}          & 1 & 10 & 1 & / & / & / & /\\				
	                                              & 2 & 25 & 6 & / & / & 10 & 1	\\	
                                                  & 3 & 65 & 16 & 20 & $-4$ & 25 & 1\\				
        	                                      & 4 & 165 & 46 & 35 & $-4$ & 60 & 1\\
                                                  & 5 & 310 & 91 & 70 & 6 & 110 & 1\\  \hline
        \multirow{5}{*}{Ammann--Beenker (8-fold)} & 1 & 16 & 1  & / & / & / & /\\
	                                              & 2 & 40 & 9 &  / & / & 16 & 1 \\ 
	                                              & 3 & 64 & 17 & / & / & 24 & 1\\ 
	                                              & 4 & 104 & 25 & 40 & $-15$ & 40 & 1\\ 
	                                              & 5 & 200 & 57 & 64 & $-23$ & 72 & 1\\	\hline
       \multirow{5}{*}{Stampfli (12-fold)}        & 1 & 24 & 1 & / & / & / & /\\	
                                                  & 2 & 60 & 13 & / & / & 24 & 1\\	
                                        	      & 3 & 120 & 25 & 36 & 0 & 48 & 1\\				
	                                              & 4 & 216 & 49 & 60 & 12 & 84 & 1\\
	                                              & 5 & 336 & 73 & 108 & 12 & 132 & 1\\
    \end{tabular}
    \caption{\textbf{The change in the number of holes $d_{\text{hole}}$ of the deployable Penrose, Ammann--Beenker, and Stampfli patterns produced by the three different design methods.}}
    \label{table:holes}
\end{table*}

Note that the expansion method transforms a closed and compact pattern into a pattern with multiple holes throughout deployment, and hence we always have $d_{\text{hole}}>0$. Moreover, by the construction of the expansion tiles, $d_{\text{hole}}$ increases strictly with the number of tiles $n$. In particular, we find that $d_{\text{hole}}$ increases linearly with $n$, and the slope is approximately $1/4$. The removal method transforms a pattern with holes into a pattern with holes throughout deployment, and different quasicrystal patterns can have highly different $d_{\text{hole}}$. In particular, some holes may merge throughout the deployment process, thereby leading to a negative $d_{\text{hole}}$. By contrast, the Hamiltonian method transforms a closed and compact pattern into a single loop throughout deployment and hence we always have $d_{\text{hole}} = 1$ regardless of the pattern size.

Below, we perform a more detailed analysis of the scaling of $d_{\text{hole}}$ with $n$ for the expansion method. To simplify our analysis, we focus on the Penrose tilings, for which all original tiles are quadrilateral. For any deployable version of them obtained by the expansion method, let $n_o$ and $n_e$ be the number of original tiles and the number of expansion tiles respectively. The total number of tiles in the deployable pattern is $n = n_e + n_o$. Let $n_{\text{int}}$ be the number of \emph{interior tiles} in the original tiling for which all sides of the tiles are shared with some hole. Let $n_{\text{bdy1}}, n_{\text{bdy2}}, n_{\text{bdy3}}$ be the number of \emph{boundary tiles} in the original tiling for which exactly one, two or three sides of the tiles are not shared with any hole in the resulting deployable pattern respectively. We have $n_o = n_{\text{int}} + n_{\text{bdy1}} + n_{\text{bdy2}} + n_{\text{bdy3}}$. 

Note that each expansion tile is connected to exactly two original tiles. Therefore, if we count the number of expansion tiles (with repetitions) using the above four types of original tiles, by the handshaking lemma we have 
\begin{equation}
    2n_e = 4n_{\text{int}}+4n_{\text{bdy1}}+3n_{\text{bdy2}}+2n_{\text{bdy3}},
\end{equation}
which yields
\begin{equation}
    n_e = 2n_{\text{int}}+2n_{\text{bdy1}}+\frac{3}{2}n_{\text{bdy2}}+n_{\text{bdy3}}
\end{equation}
and hence
\begin{equation}
    n = n_e + n_o = 3n_{\text{int}}+3n_{\text{bdy1}}+\frac{5}{2}n_{\text{bdy2}}+2n_{\text{bdy3}}.
\end{equation}
It is noteworthy that the ratio of $n_{\text{bdy1}}, n_{\text{bdy2}}, n_{\text{bdy3}}$ may vary as shown in the examples in Fig.~\ref{fig:SI_expansion}(a), and hence it is difficult to further simplify the above expression.

Next, we find the relation between the number of holes $d_{\text{hole}}$ and the number of tiles $n$. As shown in the 7 motifs of the deployable Penrose pattern in Fig.~\ref{fig:SI_vertex_star}(a), the holes can be surrounded by 3, 4, 5, 6 or 7 tiles. Denote the number of occurrence of the 7 motifs by $h_1, h_2, \dots, h_7$ respectively. As each hole corresponds to exactly one motif, we have 
\begin{equation}\label{eqt:SI_occurence}
d_{\text{hole}} = h_1 + h_2 + \dots + h_7. 
\end{equation}
Now, note that each interior tile with $m$ sides is always adjacent to exactly $m$ motifs. Also, the three types of boundary tiles and each corner tile are adjacent to 3, 2 and 1 motifs respectively. Therefore, if we count the number of tiles (with repetitions) using the motifs, we have
\begin{equation}
\begin{split}
    &5h_1 + 5h_2 + 6h_3 + 4h_4 + 3h_5 + 3h_6 + 7h_7 \\
    = \  &4n_{\text{int}} + 3n_{\text{bdy1}} + 2n_{\text{bdy2}}+n_{\text{bdy3}} \\
    = \ &\frac{4}{3} n - n_{\text{bdy1}} - \frac{4}{3}n_{\text{bdy2}} - \frac{5}{3}n_{\text{bdy3}}.
\end{split}
\end{equation}
If we assume that the occurrence of the motifs is approximately uniform, we have $h_1 \approx \dots \approx h_7 \approx d_{\text{hole}} / 7 $ and hence
\begin{equation}
    \frac{33}{7} d_{\text{hole}} \approx \frac{4}{3} n - n_{\text{bdy1}} - \frac{4}{3}n_{\text{bdy2}} - \frac{5}{3}n_{\text{bdy3}}, 
\end{equation}
which gives
\begin{equation}
    d_{\text{hole}} \approx \frac{28}{99} n - \frac{7}{33}n_{\text{bdy1}} - \frac{28}{99}n_{\text{bdy2}} - \frac{35}{99}n_{\text{bdy3}}.
\end{equation}
If we further assume that the occurrence of the three types of boundary tiles is approximately uniform, we have $n_{\text{bdy1}} \approx n_{\text{bdy2}} \approx n_{\text{bdy3}}  \approx n_b /3$ where $n_b$ is the number of boundary tiles in the original pattern, and hence
\begin{equation}\label{eqt:SI_dhole}
    d_{\text{hole}} \approx \frac{28}{99} n - \frac{7}{33}\frac{n_b}{3} - \frac{28}{99}\frac{n_b}{3} - \frac{35}{99}\frac{n_b}{3} =  \frac{28}{99} n - \frac{28}{99}n_b.
\end{equation}
As $n_b$ scales approximately with $\sqrt{n}$ and is much smaller than $n$, $d_{\text{hole}}/n$ should be slightly smaller than $28/99$, which agrees with the slope of approximately $1/4$ we observe from the example patterns. 

One can perform a similar analysis for the deployable Ammann--Beenker and Stampfli tilings obtained by the expansion method. For the Ammann--Beenker tilings, all tiles are also quadrilateral but the number of possible motifs is different, and hence the expressions in Eq.~\eqref{eqt:SI_occurence}--\eqref{eqt:SI_dhole} will be slightly different. For the Stampfli tilings, one has to separate each of the above types of interior and boundary tiles into two sub-types, one for the quadrilaterals and one for the triangles.  

\begin{table*}[t]
\small
    \centering
    \begin{tabular}{c|c|c|c|c|c|c|c}
        \multirow{2}{*}{\textbf{Pattern}} & \multirow{2}{*}{\textbf{\# layers}} & \multicolumn{2}{c|}{\textbf{Expansion}} & \multicolumn{2}{c|}{\textbf{Removal}} & \multicolumn{2}{c}{\textbf{Hamiltonian}}\\ \cline{3-8}
        & & \# tiles & $d_{\text{int}}$ & \# tiles & $d_{\text{int}}$ & \# tiles & $d_{\text{int}}$\\ \hline
       \multirow{5}{*}{Penrose (5-fold)}          & 1 & 10 & 8 & / & / & / & /\\				
	                                              & 2 & 25 & 13 & / & / & 10 & 8	\\	
                                                  & 3 & 65 & 33 & 20 & 18 & 25 & 23\\				
        	                                      & 4 & 165 & 73 & 35 & 33 & 60 & 58\\
                                                  & 5 & 310 & 128 & 70 & 38 & 110 & 108\\  \hline
        \multirow{5}{*}{Ammann--Beenker (8-fold)} & 1 & 16 & 14 & / & / & / & /\\
	                                              & 2 & 40 & 22 &  / & / & 16 & 14 \\ 
	                                              & 3 & 64 & 30 & / & / & 24 & 22\\ 
	                                              & 4 & 104 & 54 & 40 & 38 & 40 & 38\\ 
	                                              & 5 & 200 & 86 & 64 & 62 & 72 & 70\\	\hline
       \multirow{5}{*}{Stampfli (12-fold)}	      & 1 & 24 & 21 & / & / & /  & /\\	
                                                  & 2 & 60 & 33 &  / & / & 24 & 21\\	
                                        	      & 3 & 120 & 69 & 36 & 33 & 48 & 45\\				
	                                              & 4 & 216 & 117 & 60 & 33 & 84 & 81\\
	                                              & 5 & 336 & 189 & 108 & 57 & 132 & 129\\
    \end{tabular}
    \caption{\textbf{The total internal degrees of freedom $d_{\text{int}}$ of the deployable Penrose, Ammann--Beenker, and Stampfli patterns produced by the three different design methods.}}
    \label{table:dof}
\end{table*}

\subsection{Mechanics}
Finally, we study the mechanics of the patterns by considering their infinitesimal rigidity~\cite{guest2006stiffness,chen2020deterministic}. As described in~\cite{chen2020deterministic}, the rigidity of each tile in a kirigami pattern can be enforced by a set of edge and diagonal length constraints in the form of 
\begin{equation}
    g_{\text{length}}(\mathbf{x}_i,\mathbf{x}_j) = \|\mathbf{x}_i - \mathbf{x}_j\|^2 - d_{ij}^2 = 0,
\end{equation}
where $\mathbf{x}_i,\mathbf{x}_j$ are two vertices of a tile. The connectivity of the tiles can be enforced by a set of connectivity constraints in the form of 
\begin{equation}
    g_{\text{connectivity}_x}(\mathbf{x}_i,\mathbf{x}_j) = x_{i_1} - x_{j_1} = 0
\end{equation}
and 
\begin{equation}
g_{\text{connectivity}_y}(\mathbf{x}_i,\mathbf{x}_j) = x_{i_2} - x_{j_2} = 0,
\end{equation}
where $\mathbf{x}_i = (x_{i_1}, x_{i_2})$ and $\mathbf{x}_j = (x_{j_1}, x_{j_2})$ are two vertices of two connecting tiles. The above constraints can be used for constructing a rigidity matrix $\mathbf{A}$, which allows us to determine the range of motions associated with infinitesimal rigidity and hence the total internal degrees of freedom (DOF)~\cite{guest2006stiffness}:
\begin{equation}
d_{\text{int}} = 2|\mathcal{V}| - \text{rank}(\mathbf{A}) - 3,
\end{equation}
where $|\mathcal{V}|$ is the total number of vertices in the kirigami pattern. Here, the last term is used for removing the three global DOF of the entire pattern (two translational and one rotational).

We use the above rigidity matrix rank computation to assess the floppiness of the deployable quasicrystal patterns produced by the three design methods. Table~\ref{table:dof} records the value of $d_{\text{int}}$ for the deployable quasicrystal patterns. As the computation of $d_{\text{int}}$ is merely based on the length constraints (for the rigidity of the tiles) and the connectivity constraints (for the connectivity of the tiles), some modes detected by the rigidity matrix computation may be associated with tile overlaps (which can be considered as geometrical frustrations of the tiles under the deployment). In other words, $d_{\text{int}}$ serves as an upper bound for the number of physically realizable zero energy deployed states of the kirigami pattern. Fig.~\ref{fig:SI_F4}(d) shows the plot of $d_{\text{int}}$ for the deployable Penrose, Ammann--Beenker and Stampfli patterns produced by the three proposed methods. It can be observed that $d_{\text{int}}$ increases approximately linearly with the number of tiles $n$ for all three methods. In particular, the Hamiltonian method achieves the largest $d_{\text{int}}$ among the three methods, with $d_{\text{int}} \sim n$. 

To explain this, note that under the Hamiltonian method, the $n$ tiles form a single loop with each tile connected to exactly two other tiles. If all $n$ tiles are disconnected, the total DOF of the entire pattern is $3n$ (two translational and one rotational DOF for each tile). As we connect all tiles one by one, note that the $2(n-1)$ connectivity constraints for the first $n-1$ connections are always independent, while the last connection of the two ends of the chain of tiles may lead to some redundancy in the DOF counting. Therefore, we have $d_{\text{int}} \approx 3n - 2n = n$. 

\begin{figure*}[t!]
    \centering
    \includegraphics[width=0.8\textwidth]{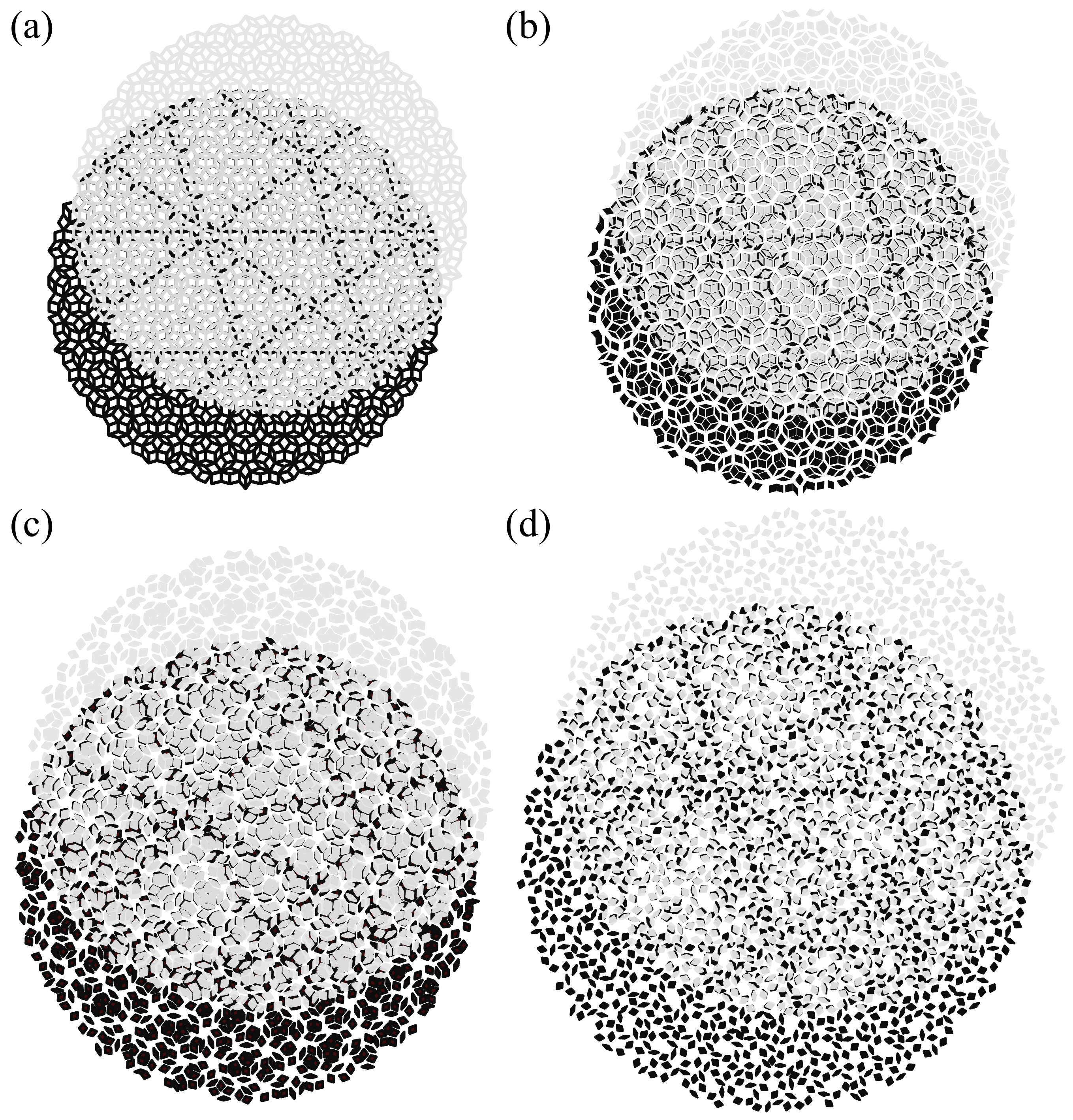}
    \caption{\textbf{Quasiperiodic translational order of deployable quasicrystal patterns}. We deploy a large Penrose kirigami pattern using the expansion tile method. At various stages of the deployment, we superimpose two differently colored copies of the configuration as shown in (a)--(d). In the resulting moir\'e pattern, the emergence of aperiodic lines indicates quasiperiodic translational order. For better visualization, in (b)--(d) the thin ideal expansion tiles are not shown while the other tiles are filled and outlined in a thickened border stroke. The patterns are not displayed to scale.}
    \label{fig:F4}
\end{figure*}

\subsection{Summary of the properties}
Our analysis of the geometrical, topological and mechanical properties of the patterns derived from the three construction methods shows they are suitable for different applications. The expansion tile method produces deployable patterns that achieve substantial size changes upon deployment without being too floppy. The tile removal method achieves deployability and shape change of the holes without much overall pattern size change. Finally, the Hamiltonian cycle method can be used to generate large pattern size change during deployment.

Also, note that for all three construction methods, the resulting deployable patterns are significantly different from periodic tilings. For instance, for the periodic rotating squares tilings with $n \times n$ tiles, one can consider the convex hull of the fully deployed configuration and easily see that the size change ratio is 
\begin{equation}
    r_s = (n^2+(n-1)^2+2(n-1))/n^2 = 2 - 1/n^2.
\end{equation}
The perimeter change ratio for the periodic rotating squares tilings is 
\begin{equation}
r_p = (8(n-1)+4)/(4n) = 2 - 1/n, 
\end{equation}
and there is a single DOF regardless of $n$. Similarly, one can see that for other periodic tilings such as the kagome (triangle-based) tilings and the hexagon tilings, the size change and perimeter change are not significantly affected by $n$. By contrast, we see that for all three construction methods proposed in this work, the size change ratio increases approximately linearly with $n$. The perimeter change ratio also increases approximately linearly with $n$ for the expansion method and the Hamiltonian method, and the internal DOF $d_{int}$ increases with $n$ for all three methods. These properties of our construction methods allow us to easily control the size and DOF of the structures by simply increasing or reducing the number of cuts and achieve different desired effects.

\begin{figure*}[t!]
    \centering
    \includegraphics[width=0.9\textwidth]{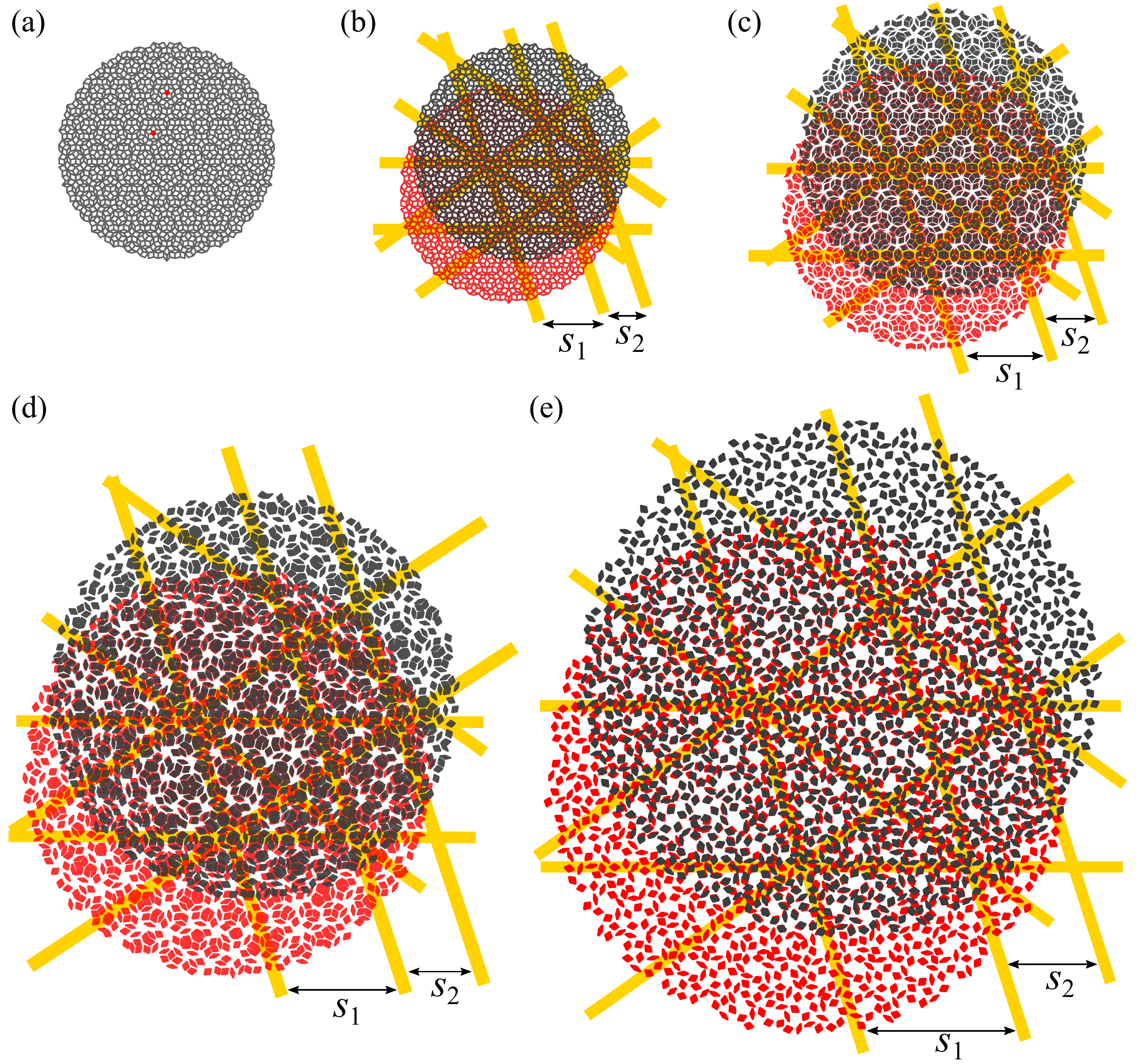}
    \caption{\textbf{An alternative visualization of the quasiperiodic translational order of deployable quasicrystal patterns}. (a)~A Penrose pattern with 1550 tiles with two reference points highlighted in red. We create a deployable Penrose kirigami pattern using the expansion tile method and overlay the reference points in two copies of the pattern for the overlay in each of the subsequent plots. The inside of tiles is left unfilled to make tile boundaries visible. (b)--(e) An alternative visualization of Fig.~\ref{fig:F4}, with the two copies of the pattern colored in red and black respectively. The observed aperiodic lines are highlighted in yellow. For better visualization, in (c)--(e) the thin ideal expansion tiles are not shown. All patterns are displayed to scale.}
    \label{fig:SI_quasiperiodic}
\end{figure*}

\section{Quasiperiodic translational order of deployable quasicrystal patterns}
Next, we consider how deployment affects pattern structure. Since quasicrystals have quasiperiodic translational order~\cite{levine1986quasicrystals}, it is natural to ask whether this order is preserved during deployment. Here we produce a large 1550-tile Penrose pattern using the Penrose pattern inflation rules for the analysis. In the contracted quasicrystal pattern, the emergence of aperiodic lines in the moir\'e pattern, as seen in Fig.~\ref{fig:F4}(a), indicates quasiperiodic translational order~\cite{levine1986quasicrystals}. Deploying the Penrose pattern using the expansion tile method and superposing it with a translated version of itself leads to moir\'e patterns. Specifically, aperiodic lines persist as the large quasicrystal is deployed (Fig.~\ref{fig:F4}(b)--(d)). To assess the spacing ratios between the lines at different stages throughout deployment, we consider an alternative visualization of the patterns as shown in Fig.~\ref{fig:SI_quasiperiodic}. Table~\ref{table:quasiperiodic} shows the spacing between the three aperiodic lines in the ``$\backslash$'' direction emerged in each configuration. It can be observed that the ratio remains almost unchanged throughout the deployment, and the value is very close to the golden ratio $\varphi = \frac{1+\sqrt{5}}{2} \approx 1.618$.

\begin{figure}[t]
    \centering
    \includegraphics[width=0.45\textwidth]{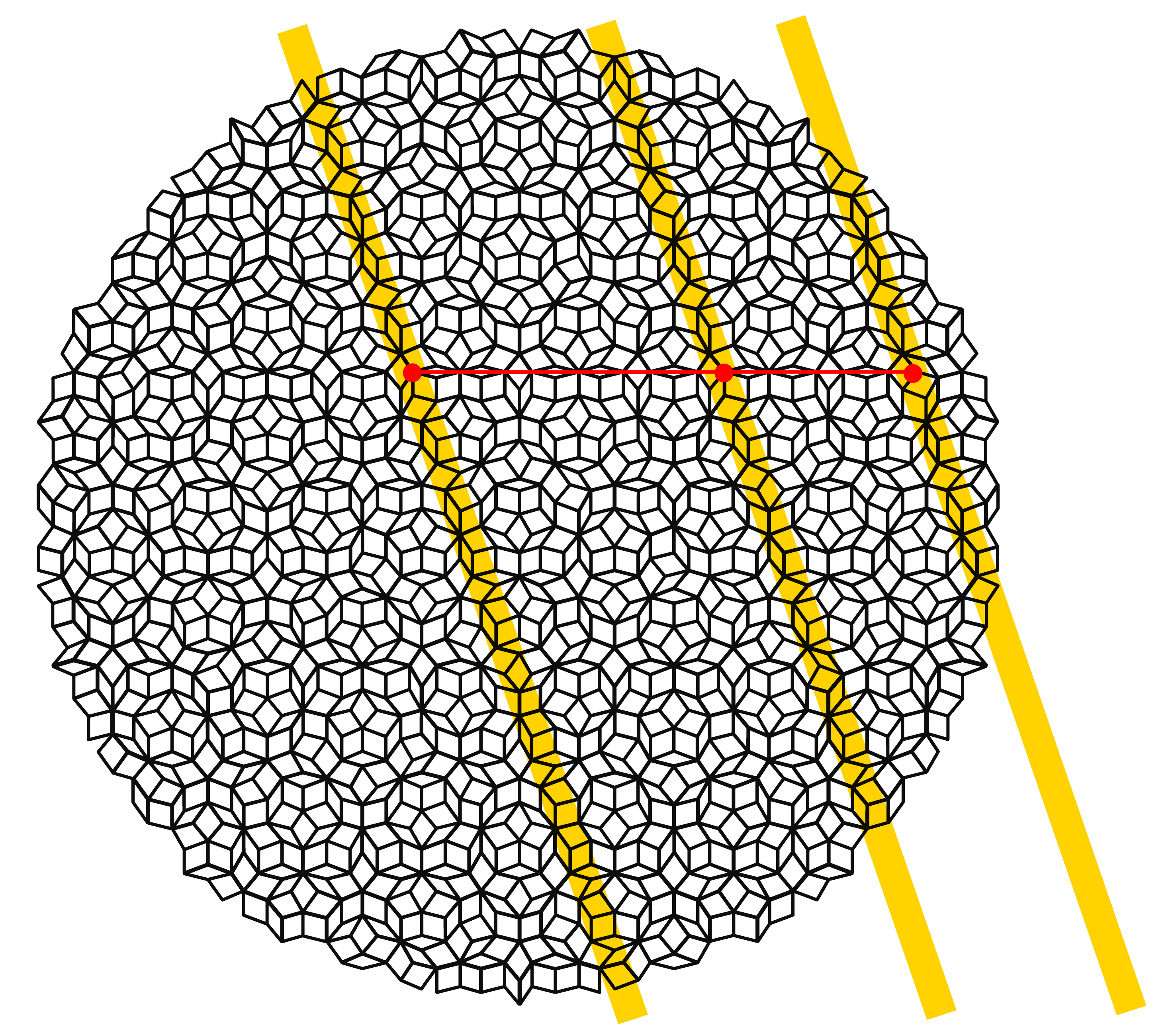}
    \caption{\textbf{The spacing between the aperiodic lines in Fig.~\ref{fig:SI_quasiperiodic}(b)}. One can measure the spacing between the lines by considering the red straight line passing through the three red dots.}
    \label{fig:SI_spacing_explanation}
\end{figure}

\begin{figure}[t]
    \centering
    \includegraphics[width=0.45\textwidth]{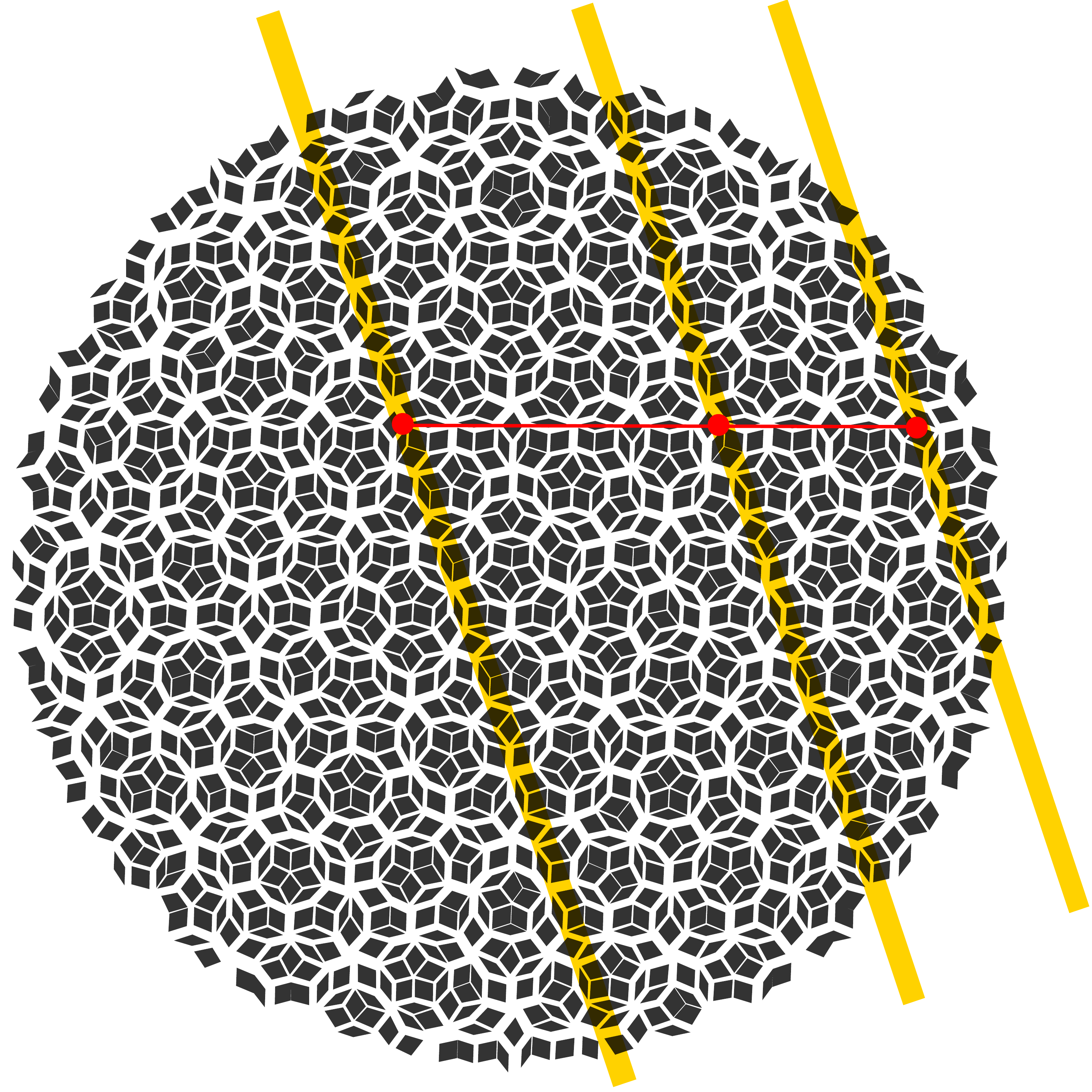}
    \caption{\textbf{The spacing between the aperiodic lines in Fig.~\ref{fig:SI_quasiperiodic}(c)}. One can measure the spacing between the lines by considering the red straight line passing through the three red dots. Note that the thin ideal expansion tiles are not shown.}
    \label{fig:SI_spacing_explanation2}
\end{figure}

\begin{table}[t!]
\small
    \centering
    \begin{tabular}{c|c|c|c}
        {\textbf{Pattern}} & {\textbf{Spacing} $s_{1}$} & {\textbf{Spacing} $s_{2}$} & {\textbf{Ratio} $s_{1} / s_{2}$}\\ \hline
       Fig.~\ref{fig:SI_quasiperiodic}(b)          & 28.2 & 17.4 & 1.62\\ 
       Fig.~\ref{fig:SI_quasiperiodic}(c)          & 37.3 & 23.0 & 1.62\\ 
       Fig.~\ref{fig:SI_quasiperiodic}(d)          & 52.5 & 32.7 & 1.61\\ 
       Fig.~\ref{fig:SI_quasiperiodic}(e)          & 68.5 & 42.9 & 1.60\\ 
    \end{tabular}
    \caption{\textbf{The spacing between the three aperiodic lines in the ``$\backslash$'' direction emerged in each overlaid image in Fig.~\ref{fig:SI_quasiperiodic}.} Here, $s_{1}$ denotes the spacing between the left line and the middle line, and $s_{2}$ denotes the spacing between the middle line and the right line. The measurement is done in the vector graphics software Inkscape.}
    \label{table:quasiperiodic}
\end{table}

To explain the relationship between the observed ratio and the golden ratio, note that the angles of every thin rhombus in the Penrose tiling are $36^{\circ}$, $144^{\circ}$, $36^{\circ}$, $144^{\circ}$, and the angles of every thick rhombus are $72^{\circ}$, $108^{\circ}$, $72^{\circ}$, $108^{\circ}$. Therefore, if the side length of the rhombi is $s$, by trigonometry one can show that the length of the longer diagonal of every thin rhombus is $a = \frac{\sqrt{10+2\sqrt{5}}}{2}s$ and the length of the shorter diagonal of every thick rhombus is $b = \frac{\sqrt{10-2\sqrt{5}}}{2}s$. Now, consider the spacing between the lines in the overlaid contracted patterns in Fig.~\ref{fig:SI_quasiperiodic}(b). As shown in Fig.~\ref{fig:SI_spacing_explanation}, $s_1$ can be measured by considering the line segment between the first two red dots, which passes through 5 thin rhombi along their longer diagonal and 3 thick rhombi along their shorter diagonal. Similarly, $s_2$ can be measured by considering the line segment between the second and third red dots, which passes through 3 thin rhombi along their longer diagonal and 2 thick rhombi along their shorter diagonal. Hence, we have
\begin{equation}
\begin{split}
    \frac{s_1}{s_2} = \frac{5a+3b}{3a+2b} &= \frac{5\frac{\sqrt{10+2\sqrt{5}}}{2}s+3\frac{\sqrt{10-2\sqrt{5}}}{2}s}{3\frac{\sqrt{10+2\sqrt{5}}}{2}s+2\frac{\sqrt{10-2\sqrt{5}}}{2}s}\\
    &= \frac{5\sqrt{\sqrt{5}+1}+3\sqrt{\sqrt{5}-1}}{3\sqrt{\sqrt{5}+1}+2\sqrt{\sqrt{5}-1}} \\
    &=\frac{1+\sqrt{5}}{2} = \varphi.
\end{split}
\end{equation}

Similarly, for the lines in the overlaid deployed patterns in Fig.~\ref{fig:SI_quasiperiodic}(c), one can see from Fig.~\ref{fig:SI_spacing_explanation2} that the first line segment passes through approximately 5 thin rhombi along their longer diagonal, 3 thick rhombi along their shorter diagonal, and 8 approximately equal gaps (each with width $c$), while the second line segment passes through approximately 3 thin rhombi along their longer diagonal, 2 thick rhombi along their shorter diagonal, and 5 approximately equal gaps (each with width $c$). Hence, we have
\begin{equation}
\begin{split}
    \frac{s_1}{s_2} \approx \frac{5a+3b+8c}{3a+2b+5c} &=  \frac{\varphi(3a+2b)+8c}{3a+2b+5c}\\
    &= \varphi + \frac{8c-5\varphi c}{3a+2b+5c}\\
    &= \varphi - \frac{5\varphi-8}{\frac{3a+2b}{c}+5}.
\end{split}
\end{equation}
Now, since $5\varphi -8 \approx 5\times 1.618 -8 = 0.09$ and $3a+2b \gg c$, we have $\frac{5\varphi -8}{\frac{3a+2b}{c}+5} \approx 0$ and hence $\frac{s_1}{s_2} \approx \varphi$. 

For the two other deployed states in Fig.~\ref{fig:SI_quasiperiodic}(d)--(e), one can assess the spacing ratio by considering the two line segments analogously. Note that the tiles may have been rotated by an approximately equal angle $\theta$ so that the line segments do not exactly pass through the diagonals of them. The deviation in the gap widths in the line segments also becomes larger. Nevertheless, we can approximate the length of the two line segments by $s_1 \approx 5\tilde{a}+3\tilde{b}+8\tilde{c}$ and $s_2 \approx 3\tilde{a}+2\tilde{b}+5\tilde{c}$, where $\tilde{a} = a \cos \theta$, $\tilde{b} = b \cos \theta$, and $\tilde{c}$ is the average gap width. Then we have
\begin{equation}
\begin{split}
    \frac{s_1}{s_2} \approx \frac{5\tilde{a}+3\tilde{b}+8\tilde{c}}{3\tilde{a}+2\tilde{b}+5\tilde{c}} &= \frac{5a+3b+8\frac{\tilde{c}}{\cos \theta}}{3a+2b+5\frac{\tilde{c}}{\cos \theta}} \\
    &=  \varphi - \frac{5\varphi-8}{\frac{3a+2b}{\tilde{c}}\cos \theta +5}.
\end{split}
\end{equation}
Again, one can see that the last term in the above expression is very small and hence $ \frac{s_1}{s_2} \approx \varphi$. This shows that the spacing ratio remains very close to the golden ratio throughout the deployment.

Altogether, the persistence of these aperiodic lines and the invariance of the spacing ratios between the lines at different stages throughout deployment demonstrate that quasiperiodic translational order is largely preserved. This highly unusual behavior makes quasicrystals a special candidate for kirigami design.

\section{Fourier transform of deployable quasicrystal patterns} \label{sect:SI_fourier}

\begin{figure*}[t!]
    \centering
    \includegraphics[width=0.8\textwidth]{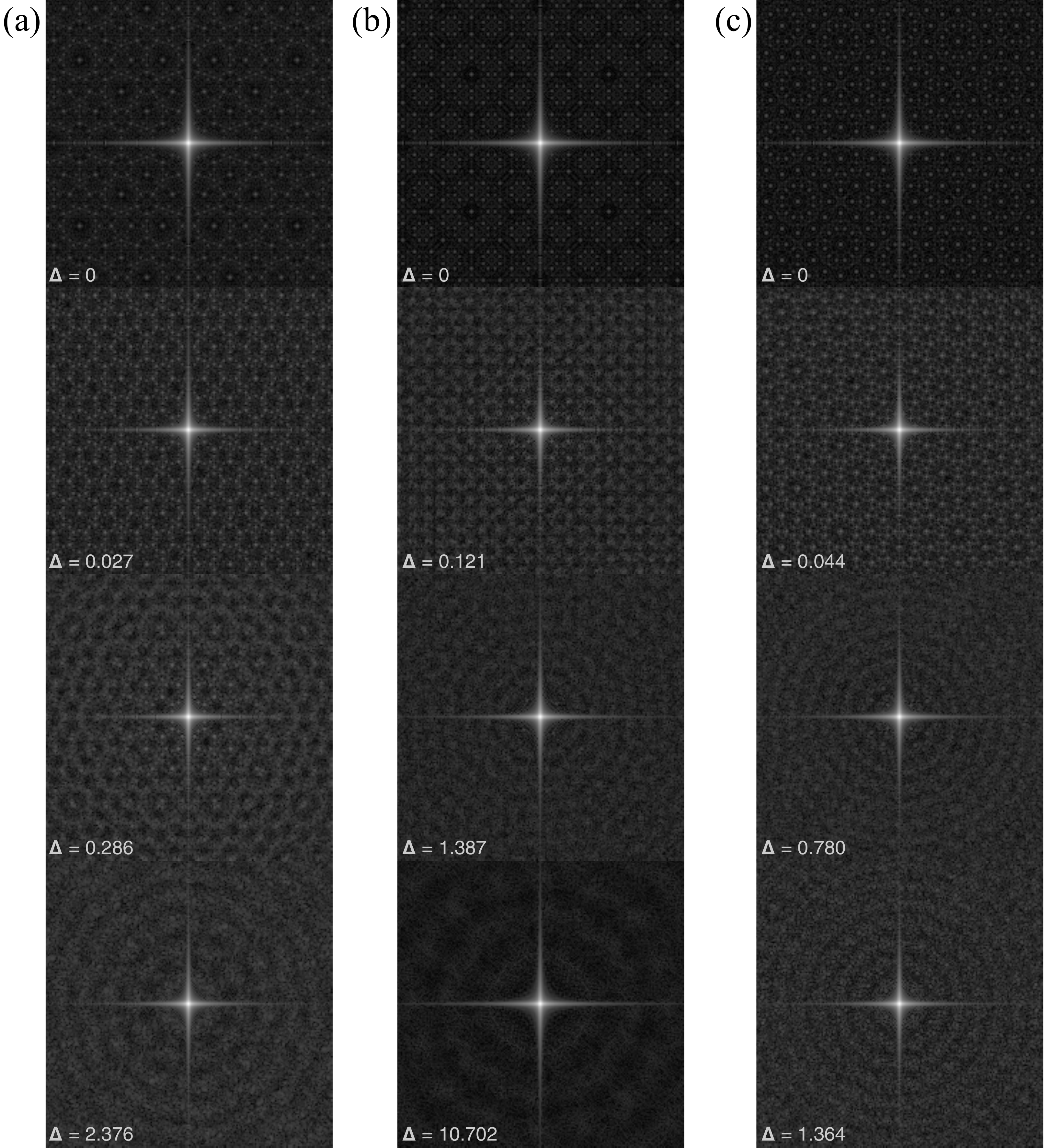}
    \caption{\textbf{Fourier transforms of deployable quasicrystal patterns}. (a) Fourier transforms of the 5-layer deployable Penrose pattern designed using the expansion method. (b) Fourier transforms of the 5-layer deployable Ammann-Beenker pattern designed using the Hamiltonian method. (c) Fourier transforms of the 5-layer deployable Stampfli pattern designed using the removal method. Each image comes from fitting the current tile vertex coordinates into a $1001 \times 1001$ image, then taking the Fourier transform as described in Section~\ref{sect:SI_fourier}. For each $1001 \times 1001$ Fourier transform image, only the center $201 \times 201$ pixels are shown so that details and symmetries can be seen. For each plot, the top row is the Fourier transform of the pattern's contracted state. The second, third, and fourth rows are, respectively, the Fourier transforms of the pattern 0.1, 1.0, and 10 seconds into deployment (simulated as described in Appendix~\ref{sect:SI_simulation}). We note that different methods need different amounts of time to deploy using our simulation methods. Since deployment speed is not uniform, we label the Fourier transform images with $\Delta$, a measure of pattern displacement. For each pattern, we calculate $\Delta(t) = \frac{1}{s}\sqrt{\frac{\sum_i^n ||\hat{v}_i(t) - \hat{v}_i(0)||_2^2}{n}}$, where $s$ is the edge length of tiles in the pattern, $n$ is the total number of tile vertices, and $\hat{v}_i(t)$ is the coordinate location of vertex $i$ at time $t$. Intuitively, $\Delta$ is a measure of displacement from initial state per tile vertex, normalized by the length of tile edges in the pattern. We can see that the largest displacement occurs with the Hamiltonian method. We remark that the bright cross around the origin in each image is an artifact caused by the Hanning window, which is applied for noise removal.}
    \label{fig:SI_fourier_images}
\end{figure*}

A different measure of order in structures is given by the Fourier transforms of their diffraction pattern; for quasicrystals, this goes back all the way back to their discovery~\cite{shechtman1984metallic,levine1984quasicrystals}. Specifically, quasicrystals have stunning structures which are ordered, but aperiodic. This is reflected in their Fourier transforms, which exhibit bright peaks with ``forbidden" orders of rotational symmetry~\cite{baake2002guide}. To study how the Fourier transform of a deployable quasicrystal pattern evolves throughout deployment, the vertex coordinates of the pattern tiles were saved at regular intervals during the deployment simulation, then fit to size $n \times n$ grayscale image arrays, where array entries corresponding to a vertex point were set to black $(0)$ while all other entries were set to white $(255)$. 

We obtained viewable FFT images by applying a Hanning window to the array, taking the Fourier transform, shifting the zero-frequency component to the center of the spectrum using the Python function \texttt{numpy.fft.fftshift}, and then taking $\log(1 + |z|)$ for each resulting complex entry $z$. Examples of these images shown in Fig.~\ref{fig:SI_fourier_images} illustrate how as deployment proceeds, quasicrystalline order is lost and the Fourier transform loses its sharp peaks of brightness. In particular, we observe that the FFT changes the most right after deployment begins, while the difference between consecutive frames in the latter stage of the deployment is less significant. These results suggest that even a small change caused by the deployment is enough to disrupt the quasicrystalline symmetry, causing a significant change in the Fourier transform relative to the smaller changes observed later on in deployment. 

Overall, for our quasicrystal kirigami patterns produced using the three construction methods, deployment breaks the mirror symmetry of the original pattern and hence the full point group symmetry of the original tiling is not preserved. Nevertheless, the 5, 8, and 12-fold symmetry can still be observed in the Fourier transforms of all snapshots as the quasicrystal kirigami patterns designed using our methods preserve rotational symmetry.

\section{Discussion}
All together, our strategies for designing a deployable quasicrystal kirigami structure are based on taking any planar tiling pattern and adding tiles to it, removing tiles from it, or changing the connectivity of tiles in it. These approaches preserve some of the symmetries of the tilings and exhibit highly unusual geometrical, topological and mechanical features throughout deployment. 

Besides the 5-fold Penrose tiling, 8-fold Ammann--Beenker tiling and 12-fold Stampfli tilings, we note that our three design methods can be applied to many other tilings with a single center of $N$-fold rotational symmetry for achieving a variety of rotational symmetry-preserving deployable structures. Since the deployable structures produced by our three proposed design methods are significantly different from traditional kirigami structures in the sense that they are ordered but aperiodic throughout the entire deployment process, they naturally complement prior kirigami approaches and may well pave the way for the design of novel deployable structures. 

The special nature of quasicrystals may also make our designed patterns useful for applications that require both order and aperiodicity. Specifically, the combination of deployability and quasiperiodic translational order suggests that the designed quasicrystal kirigami patterns may be useful for information storage and retrieval, analogous to the recent use of the Penrose tilings for visual secret sharing~\cite{yan2020penrose}, physical cryptography and unusual mechanical and optical properties given the nature and form of the deployed patterns.

\section*{Data availability}
The deployment simulation codes and the kirigami patterns are available on GitHub at \url{https://github.com/lliu12/kirigami_sim/}.

\section*{Acknowledgments}
We thank the NSF Grant No.~DMS-2002103 (to G.P.T.C.), DMR 20-11754 (to L.M.), DMREF 19-22321 (to L.M.), and EFRI 18-30901 (to L.M.) for support. \\ 


\appendix

\section{Physical model}\label{sect:SI_physical}
To verify that our deployment simulation matches how a physical system might behave, we constructed physical models of examples of the patterns and methods we used. We produced models for the Penrose pattern with the expansion method (Fig.~\ref{fig:F1}(a)), the Stampfli pattern with the removal method (Fig.~\ref{fig:F2}(d)), and the Ammann--Beenker pattern with the Hamiltonian method (Fig.~\ref{fig:F3}(d)). In the models, we used rigid cardstock for the tiles and thread for the hinges between tiles. Because the holes the thread passed through had to be within the interior of each tile, tiles in the model were not constrained together at exactly their vertices. However, the model was still able to approximate the deployment behavior of the patterns. Animations of physical models deploying were produced via stop-motion with a series of photographs of the models. Between each pair of consecutive photographs, tiles were individually moved outward. 

\section{Deployment simulation}\label{sect:SI_simulation}
The deployment simulation is implemented in Python using the Pymunk library~\cite{pymunk}, which is a wrapper for the Chipmunk2D rigid body physics engine~\cite{chipmunk}. 

More specifically, we first read the vertex coordinates of each tile of a kirigami pattern in Python and set each tile as a rigid body using the \texttt{Body} class. Each body is associated with a \texttt{Shape} defined by the vertices of its tile, which Pymunk uses to detect collisions between shapes and to ensure that the geometry of a shape is preserved throughout the simulation. We use \texttt{PinJoint} to add connection pins that link tiles as specified in the connectivity of the pattern. When a \texttt{Pinjoint} is added between two points $\mathbf{x}_i$ and $\mathbf{x}_j$ on two tiles' shapes, those points are set to remain a distance of $d_{ij}$ apart, where $d_{ij} = \|\mathbf{x}_i(0) - \mathbf{x}_j(0)\|_2$ is their Euclidean distance in the initial contracted state. If we add a \texttt{Pinjoint} between two vertices that share a position in the contracted state, those vertices will be fixed at a distance of $0$ apart. This helps preserve the connectivity of the tiles throughout the deployment. To simulate ideal expansion cuts, we set \texttt{Pinjoints} between vertices that are on opposite ends of the same edge, so they are fixed to remain a distance of that edge length apart.

To deploy the pattern, we add a spring to the simulation for every tile in the hull of the pattern. One end of the spring is attached to the center of the hull tile, and the other end is extended radially from the pattern center and attached to a circle also centered at the pattern center, but with radius larger than the maximum radius of the deployed state of the pattern (i.e. the circle is large enough such that the deployed state of the pattern can never reach it). Once the simulation starts, the springs pull all hull tiles outward to deploy the pattern until its area becomes stable (i.e. reaching the fully deployed state). This gives us the deployment path $\mathbf{x}_i(t)$ for each vertex $\mathbf{x}_i$ of each tile of the pattern.


\begin{thebibliography}{43}%
\makeatletter
\providecommand \@ifxundefined [1]{%
 \@ifx{#1\undefined}
}%
\providecommand \@ifnum [1]{%
 \ifnum #1\expandafter \@firstoftwo
 \else \expandafter \@secondoftwo
 \fi
}%
\providecommand \@ifx [1]{%
 \ifx #1\expandafter \@firstoftwo
 \else \expandafter \@secondoftwo
 \fi
}%
\providecommand \natexlab [1]{#1}%
\providecommand \enquote  [1]{``#1''}%
\providecommand \bibnamefont  [1]{#1}%
\providecommand \bibfnamefont [1]{#1}%
\providecommand \citenamefont [1]{#1}%
\providecommand \href@noop [0]{\@secondoftwo}%
\providecommand \href [0]{\begingroup \@sanitize@url \@href}%
\providecommand \@href[1]{\@@startlink{#1}\@@href}%
\providecommand \@@href[1]{\endgroup#1\@@endlink}%
\providecommand \@sanitize@url [0]{\catcode `\\12\catcode `\$12\catcode
  `\&12\catcode `\#12\catcode `\^12\catcode `\_12\catcode `\%12\relax}%
\providecommand \@@startlink[1]{}%
\providecommand \@@endlink[0]{}%
\providecommand \url  [0]{\begingroup\@sanitize@url \@url }%
\providecommand \@url [1]{\endgroup\@href {#1}{\urlprefix }}%
\providecommand \urlprefix  [0]{URL }%
\providecommand \Eprint [0]{\href }%
\providecommand \doibase [0]{http://dx.doi.org/}%
\providecommand \selectlanguage [0]{\@gobble}%
\providecommand \bibinfo  [0]{\@secondoftwo}%
\providecommand \bibfield  [0]{\@secondoftwo}%
\providecommand \translation [1]{[#1]}%
\providecommand \BibitemOpen [0]{}%
\providecommand \bibitemStop [0]{}%
\providecommand \bibitemNoStop [0]{.\EOS\space}%
\providecommand \EOS [0]{\spacefactor3000\relax}%
\providecommand \BibitemShut  [1]{\csname bibitem#1\endcsname}%
\let\auto@bib@innerbib\@empty
\bibitem [{\citenamefont {Tang}\ and\ \citenamefont
  {Yin}(2017)}]{tang2017design}%
  \BibitemOpen
  \bibfield  {author} {\bibinfo {author} {\bibfnamefont {Y.}~\bibnamefont
  {Tang}}\ and\ \bibinfo {author} {\bibfnamefont {J.}~\bibnamefont {Yin}},\
  }\bibfield  {title} {\emph {\bibinfo {title} {Design of cut unit geometry in
  hierarchical kirigami-based auxetic metamaterials for high stretchability and
  compressibility},\ }}\href@noop {} {\bibfield  {journal} {\bibinfo  {journal}
  {Extreme Mech. Lett.}\ }\textbf {\bibinfo {volume} {12}},\ \bibinfo {pages}
  {77} (\bibinfo {year} {2017})}\BibitemShut {NoStop}%
\bibitem [{\citenamefont {Blees}\ \emph {et~al.}(2015)\citenamefont {Blees},
  \citenamefont {Barnard}, \citenamefont {Rose}, \citenamefont {Roberts},
  \citenamefont {McGill}, \citenamefont {Huang}, \citenamefont {Ruyack},
  \citenamefont {Kevek}, \citenamefont {Kobrin}, \citenamefont {Muller} \emph
  {et~al.}}]{blees2015graphene}%
  \BibitemOpen
  \bibfield  {author} {\bibinfo {author} {\bibfnamefont {M.~K.}\ \bibnamefont
  {Blees}}, \bibinfo {author} {\bibfnamefont {A.~W.}\ \bibnamefont {Barnard}},
  \bibinfo {author} {\bibfnamefont {P.~A.}\ \bibnamefont {Rose}}, \bibinfo
  {author} {\bibfnamefont {S.~P.}\ \bibnamefont {Roberts}}, \bibinfo {author}
  {\bibfnamefont {K.~L.}\ \bibnamefont {McGill}}, \bibinfo {author}
  {\bibfnamefont {P.~Y.}\ \bibnamefont {Huang}}, \bibinfo {author}
  {\bibfnamefont {A.~R.}\ \bibnamefont {Ruyack}}, \bibinfo {author}
  {\bibfnamefont {J.~W.}\ \bibnamefont {Kevek}}, \bibinfo {author}
  {\bibfnamefont {B.}~\bibnamefont {Kobrin}}, \bibinfo {author} {\bibfnamefont
  {D.~A.}\ \bibnamefont {Muller}},  \emph {et~al.},\ }\bibfield  {title} {\emph
  {\bibinfo {title} {Graphene kirigami},\ }}\href@noop {} {\bibfield  {journal}
  {\bibinfo  {journal} {Nature}\ }\textbf {\bibinfo {volume} {524}},\ \bibinfo
  {pages} {204} (\bibinfo {year} {2015})}\BibitemShut {NoStop}%
\bibitem [{\citenamefont {Shyu}\ \emph {et~al.}(2015)\citenamefont {Shyu},
  \citenamefont {Damasceno}, \citenamefont {Dodd}, \citenamefont {Lamoureux},
  \citenamefont {Xu}, \citenamefont {Shlian}, \citenamefont {Shtein},
  \citenamefont {Glotzer},\ and\ \citenamefont {Kotov}}]{shyu2015kirigami}%
  \BibitemOpen
  \bibfield  {author} {\bibinfo {author} {\bibfnamefont {T.~C.}\ \bibnamefont
  {Shyu}}, \bibinfo {author} {\bibfnamefont {P.~F.}\ \bibnamefont {Damasceno}},
  \bibinfo {author} {\bibfnamefont {P.~M.}\ \bibnamefont {Dodd}}, \bibinfo
  {author} {\bibfnamefont {A.}~\bibnamefont {Lamoureux}}, \bibinfo {author}
  {\bibfnamefont {L.}~\bibnamefont {Xu}}, \bibinfo {author} {\bibfnamefont
  {M.}~\bibnamefont {Shlian}}, \bibinfo {author} {\bibfnamefont
  {M.}~\bibnamefont {Shtein}}, \bibinfo {author} {\bibfnamefont {S.~C.}\
  \bibnamefont {Glotzer}}, \ and\ \bibinfo {author} {\bibfnamefont {N.~A.}\
  \bibnamefont {Kotov}},\ }\bibfield  {title} {\emph {\bibinfo {title} {A
  kirigami approach to engineering elasticity in nanocomposites through
  patterned defects},\ }}\href@noop {} {\bibfield  {journal} {\bibinfo
  {journal} {Nat. Mater.}\ }\textbf {\bibinfo {volume} {14}},\ \bibinfo {pages}
  {785} (\bibinfo {year} {2015})}\BibitemShut {NoStop}%
\bibitem [{\citenamefont {Song}\ \emph {et~al.}(2015)\citenamefont {Song},
  \citenamefont {Wang}, \citenamefont {Lv}, \citenamefont {An}, \citenamefont
  {Liang}, \citenamefont {Ma}, \citenamefont {He}, \citenamefont {Zheng},
  \citenamefont {Huang}, \citenamefont {Yu} \emph {et~al.}}]{song2015kirigami}%
  \BibitemOpen
  \bibfield  {author} {\bibinfo {author} {\bibfnamefont {Z.}~\bibnamefont
  {Song}}, \bibinfo {author} {\bibfnamefont {X.}~\bibnamefont {Wang}}, \bibinfo
  {author} {\bibfnamefont {C.}~\bibnamefont {Lv}}, \bibinfo {author}
  {\bibfnamefont {Y.}~\bibnamefont {An}}, \bibinfo {author} {\bibfnamefont
  {M.}~\bibnamefont {Liang}}, \bibinfo {author} {\bibfnamefont
  {T.}~\bibnamefont {Ma}}, \bibinfo {author} {\bibfnamefont {D.}~\bibnamefont
  {He}}, \bibinfo {author} {\bibfnamefont {Y.-J.}\ \bibnamefont {Zheng}},
  \bibinfo {author} {\bibfnamefont {S.-Q.}\ \bibnamefont {Huang}}, \bibinfo
  {author} {\bibfnamefont {H.}~\bibnamefont {Yu}},  \emph {et~al.},\ }\bibfield
   {title} {\emph {\bibinfo {title} {Kirigami-based stretchable lithium-ion
  batteries},\ }}\href@noop {} {\bibfield  {journal} {\bibinfo  {journal} {Sci.
  Rep.}\ }\textbf {\bibinfo {volume} {5}},\ \bibinfo {pages} {1} (\bibinfo
  {year} {2015})}\BibitemShut {NoStop}%
\bibitem [{\citenamefont {Rafsanjani}\ \emph {et~al.}(2018)\citenamefont
  {Rafsanjani}, \citenamefont {Zhang}, \citenamefont {Liu}, \citenamefont
  {Rubinstein},\ and\ \citenamefont {Bertoldi}}]{rafsanjani2018kirigami}%
  \BibitemOpen
  \bibfield  {author} {\bibinfo {author} {\bibfnamefont {A.}~\bibnamefont
  {Rafsanjani}}, \bibinfo {author} {\bibfnamefont {Y.}~\bibnamefont {Zhang}},
  \bibinfo {author} {\bibfnamefont {B.}~\bibnamefont {Liu}}, \bibinfo {author}
  {\bibfnamefont {S.~M.}\ \bibnamefont {Rubinstein}}, \ and\ \bibinfo {author}
  {\bibfnamefont {K.}~\bibnamefont {Bertoldi}},\ }\bibfield  {title} {\emph
  {\bibinfo {title} {Kirigami skins make a simple soft actuator crawl},\
  }}\href@noop {} {\bibfield  {journal} {\bibinfo  {journal} {Sci. Robot.}\
  }\textbf {\bibinfo {volume} {3}},\ \bibinfo {pages} {eaar7555} (\bibinfo
  {year} {2018})}\BibitemShut {NoStop}%
\bibitem [{\citenamefont {Grima}\ and\ \citenamefont
  {Evans}(2006)}]{grima2006auxetic}%
  \BibitemOpen
  \bibfield  {author} {\bibinfo {author} {\bibfnamefont {J.~N.}\ \bibnamefont
  {Grima}}\ and\ \bibinfo {author} {\bibfnamefont {K.~E.}\ \bibnamefont
  {Evans}},\ }\bibfield  {title} {\emph {\bibinfo {title} {Auxetic behavior
  from rotating triangles},\ }}\href@noop {} {\bibfield  {journal} {\bibinfo
  {journal} {J. Mater. Sci.}\ }\textbf {\bibinfo {volume} {41}},\ \bibinfo
  {pages} {3193} (\bibinfo {year} {2006})}\BibitemShut {NoStop}%
\bibitem [{\citenamefont {Grima}\ and\ \citenamefont
  {Evans}(2000)}]{grima2000auxetic}%
  \BibitemOpen
  \bibfield  {author} {\bibinfo {author} {\bibfnamefont {J.~N.}\ \bibnamefont
  {Grima}}\ and\ \bibinfo {author} {\bibfnamefont {K.~E.}\ \bibnamefont
  {Evans}},\ }\bibfield  {title} {\emph {\bibinfo {title} {Auxetic behavior
  from rotating squares},\ }}\href@noop {} {\bibfield  {journal} {\bibinfo
  {journal} {J. Mater. Sci. Lett.}\ }\textbf {\bibinfo {volume} {19}},\
  \bibinfo {pages} {1563} (\bibinfo {year} {2000})}\BibitemShut {NoStop}%
\bibitem [{\citenamefont {Attard}\ and\ \citenamefont
  {Grima}(2008)}]{attard2008auxetic}%
  \BibitemOpen
  \bibfield  {author} {\bibinfo {author} {\bibfnamefont {D.}~\bibnamefont
  {Attard}}\ and\ \bibinfo {author} {\bibfnamefont {J.~N.}\ \bibnamefont
  {Grima}},\ }\bibfield  {title} {\emph {\bibinfo {title} {Auxetic behaviour
  from rotating rhombi},\ }}\href@noop {} {\bibfield  {journal} {\bibinfo
  {journal} {Phys. Status Solidi B}\ }\textbf {\bibinfo {volume} {245}},\
  \bibinfo {pages} {2395} (\bibinfo {year} {2008})}\BibitemShut {NoStop}%
\bibitem [{\citenamefont {Rafsanjani}\ and\ \citenamefont
  {Pasini}(2016)}]{rafsanjani2016bistable}%
  \BibitemOpen
  \bibfield  {author} {\bibinfo {author} {\bibfnamefont {A.}~\bibnamefont
  {Rafsanjani}}\ and\ \bibinfo {author} {\bibfnamefont {D.}~\bibnamefont
  {Pasini}},\ }\bibfield  {title} {\emph {\bibinfo {title} {Bistable auxetic
  mechanical metamaterials inspired by ancient geometric motifs},\ }}\href@noop
  {} {\bibfield  {journal} {\bibinfo  {journal} {Extreme Mech. Lett.}\ }\textbf
  {\bibinfo {volume} {9}},\ \bibinfo {pages} {291} (\bibinfo {year}
  {2016})}\BibitemShut {NoStop}%
\bibitem [{\citenamefont {Konakovi{\'c}}\ \emph {et~al.}(2016)\citenamefont
  {Konakovi{\'c}}, \citenamefont {Crane}, \citenamefont {Deng}, \citenamefont
  {Bouaziz}, \citenamefont {Piker},\ and\ \citenamefont
  {Pauly}}]{konakovic2016beyond}%
  \BibitemOpen
  \bibfield  {author} {\bibinfo {author} {\bibfnamefont {M.}~\bibnamefont
  {Konakovi{\'c}}}, \bibinfo {author} {\bibfnamefont {K.}~\bibnamefont
  {Crane}}, \bibinfo {author} {\bibfnamefont {B.}~\bibnamefont {Deng}},
  \bibinfo {author} {\bibfnamefont {S.}~\bibnamefont {Bouaziz}}, \bibinfo
  {author} {\bibfnamefont {D.}~\bibnamefont {Piker}}, \ and\ \bibinfo {author}
  {\bibfnamefont {M.}~\bibnamefont {Pauly}},\ }\bibfield  {title} {\emph
  {\bibinfo {title} {Beyond developable: computational design and fabrication
  with auxetic materials},\ }}\href@noop {} {\bibfield  {journal} {\bibinfo
  {journal} {ACM Trans. Graph.}\ }\textbf {\bibinfo {volume} {35}},\ \bibinfo
  {pages} {1} (\bibinfo {year} {2016})}\BibitemShut {NoStop}%
\bibitem [{\citenamefont {Celli}\ \emph {et~al.}(2018)\citenamefont {Celli},
  \citenamefont {McMahan}, \citenamefont {Ramirez}, \citenamefont {Bauhofer},
  \citenamefont {Naify}, \citenamefont {Hofmann}, \citenamefont {Audoly},\ and\
  \citenamefont {Daraio}}]{celli2018shape}%
  \BibitemOpen
  \bibfield  {author} {\bibinfo {author} {\bibfnamefont {P.}~\bibnamefont
  {Celli}}, \bibinfo {author} {\bibfnamefont {C.}~\bibnamefont {McMahan}},
  \bibinfo {author} {\bibfnamefont {B.}~\bibnamefont {Ramirez}}, \bibinfo
  {author} {\bibfnamefont {A.}~\bibnamefont {Bauhofer}}, \bibinfo {author}
  {\bibfnamefont {C.}~\bibnamefont {Naify}}, \bibinfo {author} {\bibfnamefont
  {D.}~\bibnamefont {Hofmann}}, \bibinfo {author} {\bibfnamefont
  {B.}~\bibnamefont {Audoly}}, \ and\ \bibinfo {author} {\bibfnamefont
  {C.}~\bibnamefont {Daraio}},\ }\bibfield  {title} {\emph {\bibinfo {title}
  {Shape-morphing architected sheets with non-periodic cut patterns},\
  }}\href@noop {} {\bibfield  {journal} {\bibinfo  {journal} {Soft Matter}\
  }\textbf {\bibinfo {volume} {14}},\ \bibinfo {pages} {9744} (\bibinfo {year}
  {2018})}\BibitemShut {NoStop}%
\bibitem [{\citenamefont {Konakovi{\'c}-Lukovi{\'c}}\ \emph
  {et~al.}(2018)\citenamefont {Konakovi{\'c}-Lukovi{\'c}}, \citenamefont
  {Panetta}, \citenamefont {Crane},\ and\ \citenamefont
  {Pauly}}]{konakovic2018rapid}%
  \BibitemOpen
  \bibfield  {author} {\bibinfo {author} {\bibfnamefont {M.}~\bibnamefont
  {Konakovi{\'c}-Lukovi{\'c}}}, \bibinfo {author} {\bibfnamefont
  {J.}~\bibnamefont {Panetta}}, \bibinfo {author} {\bibfnamefont
  {K.}~\bibnamefont {Crane}}, \ and\ \bibinfo {author} {\bibfnamefont
  {M.}~\bibnamefont {Pauly}},\ }\bibfield  {title} {\emph {\bibinfo {title}
  {Rapid deployment of curved surfaces via programmable auxetics},\
  }}\href@noop {} {\bibfield  {journal} {\bibinfo  {journal} {ACM Trans.
  Graph.}\ }\textbf {\bibinfo {volume} {37}},\ \bibinfo {pages} {1} (\bibinfo
  {year} {2018})}\BibitemShut {NoStop}%
\bibitem [{\citenamefont {Choi}\ \emph {et~al.}(2019)\citenamefont {Choi},
  \citenamefont {Dudte},\ and\ \citenamefont
  {Mahadevan}}]{choi2019programming}%
  \BibitemOpen
  \bibfield  {author} {\bibinfo {author} {\bibfnamefont {G.~P.~T.}\
  \bibnamefont {Choi}}, \bibinfo {author} {\bibfnamefont {L.~H.}\ \bibnamefont
  {Dudte}}, \ and\ \bibinfo {author} {\bibfnamefont {L.}~\bibnamefont
  {Mahadevan}},\ }\bibfield  {title} {\emph {\bibinfo {title} {Programming
  shape using kirigami tessellations},\ }}\href@noop {} {\bibfield  {journal}
  {\bibinfo  {journal} {Nat. Mater.}\ }\textbf {\bibinfo {volume} {18}},\
  \bibinfo {pages} {999} (\bibinfo {year} {2019})}\BibitemShut {NoStop}%
\bibitem [{\citenamefont {Choi}\ \emph {et~al.}(2021)\citenamefont {Choi},
  \citenamefont {Dudte},\ and\ \citenamefont {Mahadevan}}]{choi2020compact}%
  \BibitemOpen
  \bibfield  {author} {\bibinfo {author} {\bibfnamefont {G.~P.~T.}\
  \bibnamefont {Choi}}, \bibinfo {author} {\bibfnamefont {L.~H.}\ \bibnamefont
  {Dudte}}, \ and\ \bibinfo {author} {\bibfnamefont {L.}~\bibnamefont
  {Mahadevan}},\ }\bibfield  {title} {\emph {\bibinfo {title} {Compact
  reconfigurable kirigami},\ }}\href@noop {} {\bibfield  {journal} {\bibinfo
  {journal} {Phys. Rev. Research}\ } (\bibinfo {year} {2021})}\BibitemShut
  {NoStop}%
\bibitem [{\citenamefont {Chen}\ \emph {et~al.}(2020)\citenamefont {Chen},
  \citenamefont {Choi},\ and\ \citenamefont
  {Mahadevan}}]{chen2020deterministic}%
  \BibitemOpen
  \bibfield  {author} {\bibinfo {author} {\bibfnamefont {S.}~\bibnamefont
  {Chen}}, \bibinfo {author} {\bibfnamefont {G.~P.~T.}\ \bibnamefont {Choi}}, \
  and\ \bibinfo {author} {\bibfnamefont {L.}~\bibnamefont {Mahadevan}},\
  }\bibfield  {title} {\emph {\bibinfo {title} {Deterministic and stochastic
  control of kirigami topology},\ }}\href@noop {} {\bibfield  {journal}
  {\bibinfo  {journal} {Proc. Natl. Acad. Sci.}\ }\textbf {\bibinfo {volume}
  {117}},\ \bibinfo {pages} {4511} (\bibinfo {year} {2020})}\BibitemShut
  {NoStop}%
\bibitem [{\citenamefont {Choi}\ \emph {et~al.}(2020)\citenamefont {Choi},
  \citenamefont {Chen},\ and\ \citenamefont {Mahadevan}}]{choi2020control}%
  \BibitemOpen
  \bibfield  {author} {\bibinfo {author} {\bibfnamefont {G.~P.~T.}\
  \bibnamefont {Choi}}, \bibinfo {author} {\bibfnamefont {S.}~\bibnamefont
  {Chen}}, \ and\ \bibinfo {author} {\bibfnamefont {L.}~\bibnamefont
  {Mahadevan}},\ }\bibfield  {title} {\emph {\bibinfo {title} {Control of
  connectivity and rigidity in prismatic assemblies},\ }}\href@noop {}
  {\bibfield  {journal} {\bibinfo  {journal} {Proc. R. Soc. A}\ }\textbf
  {\bibinfo {volume} {476}},\ \bibinfo {pages} {20200485} (\bibinfo {year}
  {2020})}\BibitemShut {NoStop}%
\bibitem [{\citenamefont {Bossart}\ \emph {et~al.}(2021)\citenamefont
  {Bossart}, \citenamefont {Dykstra}, \citenamefont {van~der Laan},\ and\
  \citenamefont {Coulais}}]{bossart2021oligomodal}%
  \BibitemOpen
  \bibfield  {author} {\bibinfo {author} {\bibfnamefont {A.}~\bibnamefont
  {Bossart}}, \bibinfo {author} {\bibfnamefont {D.~M.}\ \bibnamefont
  {Dykstra}}, \bibinfo {author} {\bibfnamefont {J.}~\bibnamefont {van~der
  Laan}}, \ and\ \bibinfo {author} {\bibfnamefont {C.}~\bibnamefont
  {Coulais}},\ }\bibfield  {title} {\emph {\bibinfo {title} {Oligomodal
  metamaterials with multifunctional mechanics},\ }}\href@noop {} {\bibfield
  {journal} {\bibinfo  {journal} {Proc. Natl. Acad. Sci.}\ }\textbf {\bibinfo
  {volume} {118}},\ \bibinfo {pages} {e2018610118} (\bibinfo {year}
  {2021})}\BibitemShut {NoStop}%
\bibitem [{\citenamefont {Liu}\ \emph {et~al.}(2021)\citenamefont {Liu},
  \citenamefont {Choi},\ and\ \citenamefont {Mahadevan}}]{liu2021wallpaper}%
  \BibitemOpen
  \bibfield  {author} {\bibinfo {author} {\bibfnamefont {L.}~\bibnamefont
  {Liu}}, \bibinfo {author} {\bibfnamefont {G.~P.~T.}\ \bibnamefont {Choi}}, \
  and\ \bibinfo {author} {\bibfnamefont {L.}~\bibnamefont {Mahadevan}},\
  }\bibfield  {title} {\emph {\bibinfo {title} {Wallpaper group kirigami},\
  }}\href@noop {} {\bibfield  {journal} {\bibinfo  {journal} {Proc. R. Soc. A}\
  }\textbf {\bibinfo {volume} {477}},\ \bibinfo {pages} {20210161} (\bibinfo
  {year} {2021})}\BibitemShut {NoStop}%
\bibitem [{\citenamefont {Gr{\"u}nbaum}\ and\ \citenamefont
  {Shephard}(1986)}]{grunbaum1986tilings}%
  \BibitemOpen
  \bibfield  {author} {\bibinfo {author} {\bibfnamefont {B.}~\bibnamefont
  {Gr{\"u}nbaum}}\ and\ \bibinfo {author} {\bibfnamefont {G.~C.}\ \bibnamefont
  {Shephard}},\ }\href@noop {} {\emph {\bibinfo {title} {Tilings and
  patterns}}}\ (\bibinfo  {publisher} {WH Freeman \& Co.},\ \bibinfo {year}
  {1986})\BibitemShut {NoStop}%
\bibitem [{\citenamefont {Shechtman}\ \emph {et~al.}(1984)\citenamefont
  {Shechtman}, \citenamefont {Blech}, \citenamefont {Gratias},\ and\
  \citenamefont {Cahn}}]{shechtman1984metallic}%
  \BibitemOpen
  \bibfield  {author} {\bibinfo {author} {\bibfnamefont {D.}~\bibnamefont
  {Shechtman}}, \bibinfo {author} {\bibfnamefont {I.}~\bibnamefont {Blech}},
  \bibinfo {author} {\bibfnamefont {D.}~\bibnamefont {Gratias}}, \ and\
  \bibinfo {author} {\bibfnamefont {J.~W.}\ \bibnamefont {Cahn}},\ }\bibfield
  {title} {\emph {\bibinfo {title} {Metallic phase with long-range
  orientational order and no translational symmetry},\ }}\href@noop {}
  {\bibfield  {journal} {\bibinfo  {journal} {Phys. Rev. Lett.}\ }\textbf
  {\bibinfo {volume} {53}},\ \bibinfo {pages} {1951} (\bibinfo {year}
  {1984})}\BibitemShut {NoStop}%
\bibitem [{\citenamefont {Levine}\ and\ \citenamefont
  {Steinhardt}(1984)}]{levine1984quasicrystals}%
  \BibitemOpen
  \bibfield  {author} {\bibinfo {author} {\bibfnamefont {D.}~\bibnamefont
  {Levine}}\ and\ \bibinfo {author} {\bibfnamefont {P.~J.}\ \bibnamefont
  {Steinhardt}},\ }\bibfield  {title} {\emph {\bibinfo {title} {Quasicrystals:
  a new class of ordered structures},\ }}\href@noop {} {\bibfield  {journal}
  {\bibinfo  {journal} {Phys. Rev. Lett.}\ }\textbf {\bibinfo {volume} {53}},\
  \bibinfo {pages} {2477} (\bibinfo {year} {1984})}\BibitemShut {NoStop}%
\bibitem [{\citenamefont {Levine}\ and\ \citenamefont
  {Steinhardt}(1986)}]{levine1986quasicrystals}%
  \BibitemOpen
  \bibfield  {author} {\bibinfo {author} {\bibfnamefont {D.}~\bibnamefont
  {Levine}}\ and\ \bibinfo {author} {\bibfnamefont {P.~J.}\ \bibnamefont
  {Steinhardt}},\ }\bibfield  {title} {\emph {\bibinfo {title} {Quasicrystals.
  {I}. {D}efinition and structure},\ }}\href@noop {} {\bibfield  {journal}
  {\bibinfo  {journal} {Phys. Rev. B}\ }\textbf {\bibinfo {volume} {34}},\
  \bibinfo {pages} {596} (\bibinfo {year} {1986})}\BibitemShut {NoStop}%
\bibitem [{\citenamefont {Socolar}\ and\ \citenamefont
  {Steinhardt}(1986)}]{socolar1986quasicrystals}%
  \BibitemOpen
  \bibfield  {author} {\bibinfo {author} {\bibfnamefont {J.~E.~S.}\
  \bibnamefont {Socolar}}\ and\ \bibinfo {author} {\bibfnamefont {P.~J.}\
  \bibnamefont {Steinhardt}},\ }\bibfield  {title} {\emph {\bibinfo {title}
  {Quasicrystals. {II}. {U}nit-cell configurations},\ }}\href@noop {}
  {\bibfield  {journal} {\bibinfo  {journal} {Phys. Rev. B}\ }\textbf {\bibinfo
  {volume} {34}},\ \bibinfo {pages} {617} (\bibinfo {year} {1986})}\BibitemShut
  {NoStop}%
\bibitem [{\citenamefont {Wang}\ \emph {et~al.}(1987)\citenamefont {Wang},
  \citenamefont {Chen},\ and\ \citenamefont {Kuo}}]{wang1987two}%
  \BibitemOpen
  \bibfield  {author} {\bibinfo {author} {\bibfnamefont {N.}~\bibnamefont
  {Wang}}, \bibinfo {author} {\bibfnamefont {H.}~\bibnamefont {Chen}}, \ and\
  \bibinfo {author} {\bibfnamefont {K.}~\bibnamefont {Kuo}},\ }\bibfield
  {title} {\emph {\bibinfo {title} {Two-dimensional quasicrystal with eightfold
  rotational symmetry},\ }}\href@noop {} {\bibfield  {journal} {\bibinfo
  {journal} {Phys. Rev. Lett.}\ }\textbf {\bibinfo {volume} {59}},\ \bibinfo
  {pages} {1010} (\bibinfo {year} {1987})}\BibitemShut {NoStop}%
\bibitem [{\citenamefont {Socolar}(1989)}]{socolar1989simple}%
  \BibitemOpen
  \bibfield  {author} {\bibinfo {author} {\bibfnamefont {J.~E.~S.}\
  \bibnamefont {Socolar}},\ }\bibfield  {title} {\emph {\bibinfo {title}
  {Simple octagonal and dodecagonal quasicrystals},\ }}\href@noop {} {\bibfield
   {journal} {\bibinfo  {journal} {Phys. Rev. B}\ }\textbf {\bibinfo {volume}
  {39}},\ \bibinfo {pages} {10519} (\bibinfo {year} {1989})}\BibitemShut
  {NoStop}%
\bibitem [{\citenamefont {Baake}\ and\ \citenamefont
  {Joseph}(1990)}]{baake1990ideal}%
  \BibitemOpen
  \bibfield  {author} {\bibinfo {author} {\bibfnamefont {M.}~\bibnamefont
  {Baake}}\ and\ \bibinfo {author} {\bibfnamefont {D.}~\bibnamefont {Joseph}},\
  }\bibfield  {title} {\emph {\bibinfo {title} {Ideal and defective vertex
  configurations in the planar octagonal quasilattice},\ }}\href@noop {}
  {\bibfield  {journal} {\bibinfo  {journal} {Phys. Rev. B}\ }\textbf {\bibinfo
  {volume} {42}},\ \bibinfo {pages} {8091} (\bibinfo {year}
  {1990})}\BibitemShut {NoStop}%
\bibitem [{\citenamefont {Baake}\ \emph {et~al.}(1994)\citenamefont {Baake},
  \citenamefont {Ben-Abraham}, \citenamefont {Klitzing}, \citenamefont
  {Kramer},\ and\ \citenamefont {Schlottman}}]{baake1994classification}%
  \BibitemOpen
  \bibfield  {author} {\bibinfo {author} {\bibfnamefont {M.}~\bibnamefont
  {Baake}}, \bibinfo {author} {\bibfnamefont {S.~I.}\ \bibnamefont
  {Ben-Abraham}}, \bibinfo {author} {\bibfnamefont {R.}~\bibnamefont
  {Klitzing}}, \bibinfo {author} {\bibfnamefont {P.}~\bibnamefont {Kramer}}, \
  and\ \bibinfo {author} {\bibfnamefont {M.}~\bibnamefont {Schlottman}},\
  }\bibfield  {title} {\emph {\bibinfo {title} {Classification of local
  configurations in quasicrystals},\ }}\href@noop {} {\bibfield  {journal}
  {\bibinfo  {journal} {Acta Cryst. A}\ }\textbf {\bibinfo {volume} {50}},\
  \bibinfo {pages} {553} (\bibinfo {year} {1994})}\BibitemShut {NoStop}%
\bibitem [{\citenamefont {Senechal}(1996)}]{senechal1996quasicrystals}%
  \BibitemOpen
  \bibfield  {author} {\bibinfo {author} {\bibfnamefont {M.}~\bibnamefont
  {Senechal}},\ }\href@noop {} {\emph {\bibinfo {title} {Quasicrystals and
  geometry}}}\ (\bibinfo  {publisher} {Cambridge University Press},\ \bibinfo
  {year} {1996})\BibitemShut {NoStop}%
\bibitem [{\citenamefont {Nagaoka}\ \emph {et~al.}(2018)\citenamefont
  {Nagaoka}, \citenamefont {Zhu}, \citenamefont {Eggert},\ and\ \citenamefont
  {Chen}}]{nagaoka2018single}%
  \BibitemOpen
  \bibfield  {author} {\bibinfo {author} {\bibfnamefont {Y.}~\bibnamefont
  {Nagaoka}}, \bibinfo {author} {\bibfnamefont {H.}~\bibnamefont {Zhu}},
  \bibinfo {author} {\bibfnamefont {D.}~\bibnamefont {Eggert}}, \ and\ \bibinfo
  {author} {\bibfnamefont {O.}~\bibnamefont {Chen}},\ }\bibfield  {title}
  {\emph {\bibinfo {title} {Single-component quasicrystalline nanocrystal
  superlattices through flexible polygon tiling rule},\ }}\href@noop {}
  {\bibfield  {journal} {\bibinfo  {journal} {Science}\ }\textbf {\bibinfo
  {volume} {362}},\ \bibinfo {pages} {1396} (\bibinfo {year}
  {2018})}\BibitemShut {NoStop}%
\bibitem [{\citenamefont {Ahn}\ \emph {et~al.}(2018)\citenamefont {Ahn},
  \citenamefont {Moon}, \citenamefont {Kim}, \citenamefont {Kim}, \citenamefont
  {Shin}, \citenamefont {Kim}, \citenamefont {Cha}, \citenamefont {Kahng},
  \citenamefont {Kim}, \citenamefont {Koshino} \emph {et~al.}}]{ahn2018dirac}%
  \BibitemOpen
  \bibfield  {author} {\bibinfo {author} {\bibfnamefont {S.~J.}\ \bibnamefont
  {Ahn}}, \bibinfo {author} {\bibfnamefont {P.}~\bibnamefont {Moon}}, \bibinfo
  {author} {\bibfnamefont {T.-H.}\ \bibnamefont {Kim}}, \bibinfo {author}
  {\bibfnamefont {H.-W.}\ \bibnamefont {Kim}}, \bibinfo {author} {\bibfnamefont
  {H.-C.}\ \bibnamefont {Shin}}, \bibinfo {author} {\bibfnamefont {E.~H.}\
  \bibnamefont {Kim}}, \bibinfo {author} {\bibfnamefont {H.~W.}\ \bibnamefont
  {Cha}}, \bibinfo {author} {\bibfnamefont {S.-J.}\ \bibnamefont {Kahng}},
  \bibinfo {author} {\bibfnamefont {P.}~\bibnamefont {Kim}}, \bibinfo {author}
  {\bibfnamefont {M.}~\bibnamefont {Koshino}},  \emph {et~al.},\ }\bibfield
  {title} {\emph {\bibinfo {title} {Dirac electrons in a dodecagonal graphene
  quasicrystal},\ }}\href@noop {} {\bibfield  {journal} {\bibinfo  {journal}
  {Science}\ }\textbf {\bibinfo {volume} {361}},\ \bibinfo {pages} {782}
  (\bibinfo {year} {2018})}\BibitemShut {NoStop}%
\bibitem [{\citenamefont {Penrose}(1974)}]{penrose1974role}%
  \BibitemOpen
  \bibfield  {author} {\bibinfo {author} {\bibfnamefont {R.}~\bibnamefont
  {Penrose}},\ }\bibfield  {title} {\emph {\bibinfo {title} {The role of
  aesthetics in pure and applied mathematical research},\ }}\href@noop {}
  {\bibfield  {journal} {\bibinfo  {journal} {Bull. Inst. Math. Appl.}\
  }\textbf {\bibinfo {volume} {10}},\ \bibinfo {pages} {266} (\bibinfo {year}
  {1974})}\BibitemShut {NoStop}%
\bibitem [{\citenamefont {Ammann}\ \emph {et~al.}(1992)\citenamefont {Ammann},
  \citenamefont {Gr{\"u}nbaum},\ and\ \citenamefont
  {Shephard}}]{ammann1992aperiodic}%
  \BibitemOpen
  \bibfield  {author} {\bibinfo {author} {\bibfnamefont {R.}~\bibnamefont
  {Ammann}}, \bibinfo {author} {\bibfnamefont {B.}~\bibnamefont
  {Gr{\"u}nbaum}}, \ and\ \bibinfo {author} {\bibfnamefont {G.~C.}\
  \bibnamefont {Shephard}},\ }\bibfield  {title} {\emph {\bibinfo {title}
  {Aperiodic tiles},\ }}\href@noop {} {\bibfield  {journal} {\bibinfo
  {journal} {Discrete Comput. Geom.}\ }\textbf {\bibinfo {volume} {8}},\
  \bibinfo {pages} {1} (\bibinfo {year} {1992})}\BibitemShut {NoStop}%
\bibitem [{\citenamefont {Stampfli}(1986)}]{stampfli1986dodecagonal}%
  \BibitemOpen
  \bibfield  {author} {\bibinfo {author} {\bibfnamefont {P.}~\bibnamefont
  {Stampfli}},\ }\bibfield  {title} {\emph {\bibinfo {title} {A dodecagonal
  quasiperiodic lattice in two dimensions},\ }}\href@noop {} {\bibfield
  {journal} {\bibinfo  {journal} {Helv. Phys. Acta}\ }\textbf {\bibinfo
  {volume} {59}},\ \bibinfo {pages} {1260} (\bibinfo {year}
  {1986})}\BibitemShut {NoStop}%
\bibitem [{\citenamefont {You}\ and\ \citenamefont
  {Pellegrino}(1997)}]{you1997foldable}%
  \BibitemOpen
  \bibfield  {author} {\bibinfo {author} {\bibfnamefont {Z.}~\bibnamefont
  {You}}\ and\ \bibinfo {author} {\bibfnamefont {S.}~\bibnamefont
  {Pellegrino}},\ }\bibfield  {title} {\emph {\bibinfo {title} {Foldable bar
  structures},\ }}\href@noop {} {\bibfield  {journal} {\bibinfo  {journal}
  {Int. J. Solids Struct.}\ }\textbf {\bibinfo {volume} {34}},\ \bibinfo
  {pages} {1825} (\bibinfo {year} {1997})}\BibitemShut {NoStop}%
\bibitem [{\citenamefont {Patel}\ and\ \citenamefont
  {Ananthasuresh}(2007)}]{patel2007kinematic}%
  \BibitemOpen
  \bibfield  {author} {\bibinfo {author} {\bibfnamefont {J.}~\bibnamefont
  {Patel}}\ and\ \bibinfo {author} {\bibfnamefont {G.~K.}\ \bibnamefont
  {Ananthasuresh}},\ }\bibfield  {title} {\emph {\bibinfo {title} {A kinematic
  theory for radially foldable planar linkages},\ }}\href@noop {} {\bibfield
  {journal} {\bibinfo  {journal} {Int. J. Solids Struct.}\ }\textbf {\bibinfo
  {volume} {44}},\ \bibinfo {pages} {6279} (\bibinfo {year}
  {2007})}\BibitemShut {NoStop}%
\bibitem [{\citenamefont {Kiper}\ \emph {et~al.}(2008)\citenamefont {Kiper},
  \citenamefont {S{\"o}ylemez},\ and\ \citenamefont
  {Ki{\c{s}}isel}}]{kiper2008family}%
  \BibitemOpen
  \bibfield  {author} {\bibinfo {author} {\bibfnamefont {G.}~\bibnamefont
  {Kiper}}, \bibinfo {author} {\bibfnamefont {E.}~\bibnamefont {S{\"o}ylemez}},
  \ and\ \bibinfo {author} {\bibfnamefont {A.~{\"O}.}\ \bibnamefont
  {Ki{\c{s}}isel}},\ }\bibfield  {title} {\emph {\bibinfo {title} {A family of
  deployable polygons and polyhedra},\ }}\href@noop {} {\bibfield  {journal}
  {\bibinfo  {journal} {Mech. Mach. Theory}\ }\textbf {\bibinfo {volume}
  {43}},\ \bibinfo {pages} {627} (\bibinfo {year} {2008})}\BibitemShut
  {NoStop}%
\bibitem [{\citenamefont {Cabras}\ and\ \citenamefont
  {Brun}(2014)}]{cabras2014auxetic}%
  \BibitemOpen
  \bibfield  {author} {\bibinfo {author} {\bibfnamefont {L.}~\bibnamefont
  {Cabras}}\ and\ \bibinfo {author} {\bibfnamefont {M.}~\bibnamefont {Brun}},\
  }\bibfield  {title} {\emph {\bibinfo {title} {Auxetic two-dimensional
  lattices with {P}oisson's ratio arbitrarily close to $-1$},\ }}\href@noop {}
  {\bibfield  {journal} {\bibinfo  {journal} {Proc. R. Soc. A}\ }\textbf
  {\bibinfo {volume} {470}},\ \bibinfo {pages} {20140538} (\bibinfo {year}
  {2014})}\BibitemShut {NoStop}%
\bibitem [{supplementary()}]{supplementary}%
  \BibitemOpen
  \href@noop {} {\bibinfo {title} {See Supplemental Material at [url will be
  inserted by publisher] for the videos of the deployment of physical 
  and numerical models of several quasicrystal kirigami patterns}\ }\BibitemShut {NoStop}%
\bibitem [{\citenamefont {Tutte}(1956)}]{tutte1956theorem}%
  \BibitemOpen
  \bibfield  {author} {\bibinfo {author} {\bibfnamefont {W.~T.}\ \bibnamefont
  {Tutte}},\ }\bibfield  {title} {\emph {\bibinfo {title} {A theorem on planar
  graphs},\ }}\href@noop {} {\bibfield  {journal} {\bibinfo  {journal} {T.
  Amer. Math. Soc.}\ }\textbf {\bibinfo {volume} {82}},\ \bibinfo {pages} {99}
  (\bibinfo {year} {1956})}\BibitemShut {NoStop}%
\bibitem [{\citenamefont {Guest}(2006)}]{guest2006stiffness}%
  \BibitemOpen
  \bibfield  {author} {\bibinfo {author} {\bibfnamefont {S.}~\bibnamefont
  {Guest}},\ }\bibfield  {title} {\emph {\bibinfo {title} {The stiffness of
  prestressed frameworks: a unifying approach},\ }}\href@noop {} {\bibfield
  {journal} {\bibinfo  {journal} {Int. J.Solids Struct.}\ }\textbf {\bibinfo
  {volume} {43}},\ \bibinfo {pages} {842} (\bibinfo {year} {2006})}\BibitemShut
  {NoStop}%
\bibitem [{\citenamefont {Baake}(2002)}]{baake2002guide}%
  \BibitemOpen
  \bibfield  {author} {\bibinfo {author} {\bibfnamefont {M.}~\bibnamefont
  {Baake}},\ }in\ \href@noop {} {\emph {\bibinfo {booktitle} {Quasicrystals}}}\
  (\bibinfo  {publisher} {Springer},\ \bibinfo {year} {2002})\ pp.\ \bibinfo
  {pages} {17--48}\BibitemShut {NoStop}%
\bibitem [{\citenamefont {Yan}\ \emph {et~al.}(2020)\citenamefont {Yan},
  \citenamefont {Yan}, \citenamefont {Liu},\ and\ \citenamefont
  {Lu}}]{yan2020penrose}%
  \BibitemOpen
  \bibfield  {author} {\bibinfo {author} {\bibfnamefont {X.}~\bibnamefont
  {Yan}}, \bibinfo {author} {\bibfnamefont {W.~Q.}\ \bibnamefont {Yan}},
  \bibinfo {author} {\bibfnamefont {L.}~\bibnamefont {Liu}}, \ and\ \bibinfo
  {author} {\bibfnamefont {Y.}~\bibnamefont {Lu}},\ }\bibfield  {title} {\emph
  {\bibinfo {title} {Penrose tiling for visual secret sharing},\ }}\href@noop
  {} {\bibfield  {journal} {\bibinfo  {journal} {Multimed. Tools. Appl.}\
  }\textbf {\bibinfo {volume} {79}},\ \bibinfo {pages} {32693} (\bibinfo {year}
  {2020})}\BibitemShut {NoStop}%
\bibitem [{\citenamefont {Blomqvist}()}]{pymunk}%
  \BibitemOpen
  \bibfield  {author} {\bibinfo {author} {\bibfnamefont {V.}~\bibnamefont
  {Blomqvist}},\ }\href@noop {} {\bibinfo {title} {Pymunk},\ }\bibinfo
  {howpublished} {\url{http://www.pymunk.org}}\BibitemShut {NoStop}%
\bibitem [{chi()}]{chipmunk}%
  \BibitemOpen
  \href@noop {} {\bibinfo {title} {{Chipmunk2D}},\ }\bibinfo {howpublished}
  {\url{http://chipmunk-physics.net/}}\BibitemShut {NoStop}%
\end{thebibliography}
\end{document}